\documentclass[useAMS,usenatbib]{mn2e}
\usepackage{graphicx}

\title[Long-term monitoring of MCG-6-30-15]{Long-term monitoring of
  the archetype Seyfert galaxy MCG-6-30-15: X-ray, optical and near-IR
  variability of the corona, disc and torus.}
 
\author[Lira et al.]{P. Lira$^{1}$, P. Ar\'evalo$^{2}$, P. Uttley$^{3}$, I. M. M. McHardy$^{4}$, L. Videla$^{5}$\\
$^{1}$Departmento de Astronomia, Universidad de Chile, Camino del Observatorio 1515, Santiago, Chile\\
$^{2}$Instituto de F\'isica y Astronom\'ia, Facultad de Ciencias, Universidad de Valpara\'iso, Gran Breta\~na 1111, Valpara\'iso, Chile\\
$^{3}$Anton Pannekoek Institute, University of Amsterdam, Science Park 904, NL-1098 XH Amsterdam, The Netherlands\\
$^{4}$Department of Physics and Astronomy, The University, Southampton SO17 1BJ\\
$^{5}$Joint ALMA Observatory, Alonso de C\'ordova 3107, Vitacura, Santiago, Chile
}
\begin{document}

\date{}

\pagerange{\pageref{firstpage}--\pageref{lastpage}} \pubyear{2002}

\maketitle

\label{firstpage}

\begin{abstract}
We present long term monitoring of MCG-6-30-15 in X-rays, optical and
near-IR wavelengths, collected over five years of monitoring. We
determine the power spectrum density of all the observed bands and
show that after taking into account the host contamination similar
power is observed in the optical and near-IR bands. There is evidence
for a correlation between the light curves of the X-ray photon flux
and the optical B-band, but it is not possible to determine a lag with
certainty, with the most likely value being around zero days. Strong
correlation is seen between the optical and near-IR bands. Cross
correlation analysis shows some complex probability distributions and
lags that range from 10 to 20 days, with the near-IR following the
optical variations. Filtering the light curves in frequency space
shows that the strongest correlations are those corresponding to the
shortest time-scales. We discuss the nature of the X-ray variability
and conclude that this is intrinsic and cannot be accounted for by
absorption episodes due to material intervening in the line of
sight. It is also found that the lags agree with the relation $\tau
\propto \lambda^{4/3}$, as expected for an optically thick
geometrically thin accretion disc, although for a larger disc than
that predicted by the estimated black hole mass and accretion rate in
MCG-6-30-15. The cross correlation analysis suggests that the torus is
located at $\sim 20$ light-days from the central source and at most at
$\sim 50$ light-days from the central region. This implies an AGN
bolometric luminosity of $\sim 3 \times 10^{43}$ ergs/s/cm$^2$.

\end{abstract}
 

\section{Introduction}
 One of the defining characteristics of Active Galactic Nuclei (AGN) is
their strong variability, which is observed on time scales of seconds
to years, and over a very broad wavelength range. The often complex,
yet evident connection between the variations seen at different
wavelengths has also been firmly established. These can be used to
unveil the physical processes behind the emission and the relations
between the different regions responsible for the variations, such as
the corona (producing the X-ray emission), the accretion disc
(producing the UV, optical and likely near-IR emission) and the dusty
torus (producing near and mid-IR emission).

Multiwavelength, high quality light curves, (i.e., those with the
desired energy coverage and time sampling), are not easy to
obtain. The advent of current and future time domain surveys are
helping to overcome this at least partially, yielding well sampled
light curves for a huge number of sources, but usually limited to the
optical band. Hence, multiwavelength monitoring data are still
obtained from targeted observations of AGN from space and ground
facilities, and it will remain this way for the time being.

The emerging picture from the analysis of long term, well sampled,
multiwavelength observations of AGN is clear in its most simple
version: variation patterns can be mapped at different wavelengths and
observed lags are roughly consistent with the time it takes for
radiation to travel from the center to the edges of the system. This
confirms the expected temperature stratification, with the hot corona
encompassing only a few gravitational radii ($R_g = GM/c^2$, where M
is the black hole mass) around the black hole, the accretion disk
extending from close to the location of the corona to hundreds or
thousands of $R_g$, and the dusty torus appearing at the radii where
dust can survive the prevailing temperatures (Suganuma et al.~2006;
Ar\'evalo et al.~2009, 2009; Breedt et al.~2009, 2010; Lira et
al.~2011; Cameron et al.~2012; McHardy et al.~2014; Shappee et
al.~2013; Edelson et al.~2015). The details, however, are still
somewhat murky, which is not surprising as these are most likely
function of several parameters such as black hole mass, accretion
rate, system geometry and inclination, etc., which are bound to change
from source to source, while for a single source, the accretion rate
might change from one observation to the next.

MCG-6-30-15 is one of the best studied AGN in X-rays (Ar\'evalo et
al., 2005; McHardy et al.~2005; Miniutti et al.~2007; Miller et al.,
2008, 2009; Chiang \& Fabian 2011; Emmanoulopoulos et al.~2011; Noda
et al.~2011; Marinucci et al.~2014; Kara et al.~2014; Ludlam et
al.~2015; just to name some of the articles from the last 10
years). The interest in its X-ray emission is due to the presence of
the broad red-shifted K$\alpha$ line, indicating X-ray reflection from
the innermost region of the accretion disc (Tanaka et al.~1995), and
its high variability (as first reported by Pounds et al.~1986 and
Nandra et al.~1990), as expected from accretion onto a black hole of
only a few million solar masses (Done \& Gierlinski 2005; Ludlam et
al.~2015). Indeed, MCG-6-30-15 is usually defined as a member of the
Narrow Line Seyfert 1 (NLS1) class.

However, no monitoring campaign has aimed at longer wavelengths. In
fact, to our knowledge, no reverberation campaign has been successfully
carried out on this source and black hole mass estimates rely on the
galactic properties of its host galaxy and X-ray variability
derivations (McHardy et al.~2005 and references therein).

We wanted to remedy this situation by conducting a long monitoring
campaign of MCG-6-30-15 from the X-rays to the near-IR
wavelengths. This very broadband approach will give us an
understanding of the relation between the corona, disc and torus
components in this AGN. This paper is organised as follows: Section 2
presents the data acquisition and reduction; Section 3 presents the
results on the determination of the Power Spectral Density, the Cross
Correlation analysis and the frequency filtered light curves; Section
4 presents a discussion on the nature of the X-ray emission, the
near-IR emission from the disc and torus and an upper limit to the
bolometric luminosity of MCG-6-30-15; Section 5 presents the
summary. Throughout this work we assume a distance to MCG-6-30-15 of 37
Mpc (z=0.008, Ho=65 km/s/Mpc).

\section{Data acquisition and analysis}

\begin{figure*}
\centering
\includegraphics[scale=1.2,angle=0,trim=100 0 0 0]{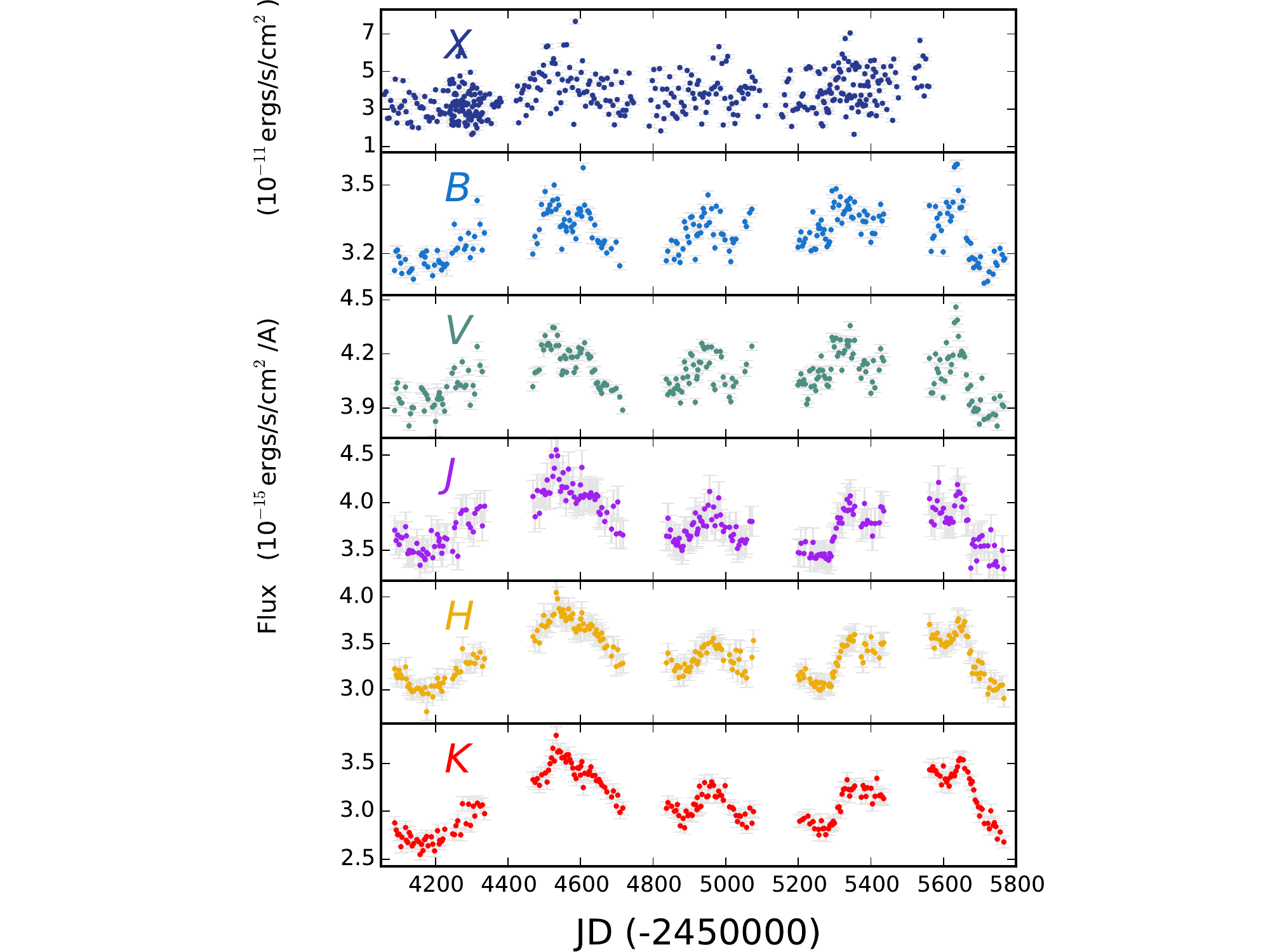}
\caption{Observed X-ray, optical and near-IR light curves for
  MCG-6-30-15. Observations have not been corrected for host
  contribution or intrinsic extinction.}
\label{mcg63015lcs}
\end{figure*}

We monitored MCG-6-30-15 in the X-ray band with the {\em Rossi X-ray
  Timing Explorer (RXTE)} taking 1 kilo-second exposure snapshots
typically every 4 days, but including some intensive periods with
snapshots every 1 or 2 days. In the present paper, we include the
X-ray data that are contemporaneous to our ground based observations
(see below). The X-ray data were obtained using the {\em RXTE}
Proportional Counter Array (PCA), which is sensitive in the range 3 to
60 keV and consists of five Proportional Counter Units (PCUs). We only
extracted data from PCU 2. We use standard good-time interval
selection criteria. Background data were created using the combined
faint background model and the SAA history. We extracted spectra in
the 3-–12 keV energy range for each snapshot and used XSPEC to fit a
simple absorbed power-law model with the absorption column fixed at
the Galactic value, $N_H = 4 \times 10^{20}$ cm$^{-2}$ to obtain an
estimate of the 2–-10 keV flux and its error. For further details on
data reduction of RXTE observations see Ar\'evalo et al.~(2008).

B, V, J, H, and K observations were obtained between August 2006 and
July 2011 with the ANDICAM camera mounted on the 1.3m telescope at
CTIO and operated by the SMARTS consortium. ANDICAM allows
simultaneous observations in the optical and near-IR by using a
dichroic with a CCD and a HgCdTe array. A movable mirror allows
dithering in the IR while an optical exposure remains still. The
average sampling of the light curves was 4.5 days. 

The optical data reduction followed the usual steps of bias
subtraction and flatfielding. The images were convolved to match the
point spread function of the worst acceptable seeing. Relative
photometry of MCG-6-30-15 and nearby stars was achieved by obtaining
aperture photometry with a radius of 1.7 arcseconds. The final flux
calibration was determined by observing standard photometric stars
during photometric conditions. The data were corrected for Galactic
foreground absorption of $A_V=0.165$. For more details on the optical
data reduction see Ar\'evalo et al.~(2009).

The near-IR data reduction followed the standard steps of dark
subtraction, flat fielding, and sky subtraction using consecutive
jittered frames. The light curves were constructed from relative
photometry obtained through a fixed aperture with diameter 2.74 arcsec
after all images were taken to a common seeing. Flux calibration was
obtained using the computed 2MASS magnitudes of the comparison
stars. Photometric errors were obtained as the squared sum of the
standard deviation due to the Poissonian noise of the source-plus-sky
flux within the aperture, plus the uncertainty due the measurement
itself. This last error was estimated as the standard deviation in the
photometry of stars available in the field of view in consecutive
exposures. For more details see Lira et al.~(2011).

Light curves in the X-ray, B, V, J, H, and K bands for MCG-6-30-15 are
presented in Figure \ref{mcg63015lcs}. 

\section{Results}

\subsection{The Power Spectral Densities}

\begin{figure}
\centering
\includegraphics[scale=0.55,angle=0,trim=30 0 0 0]{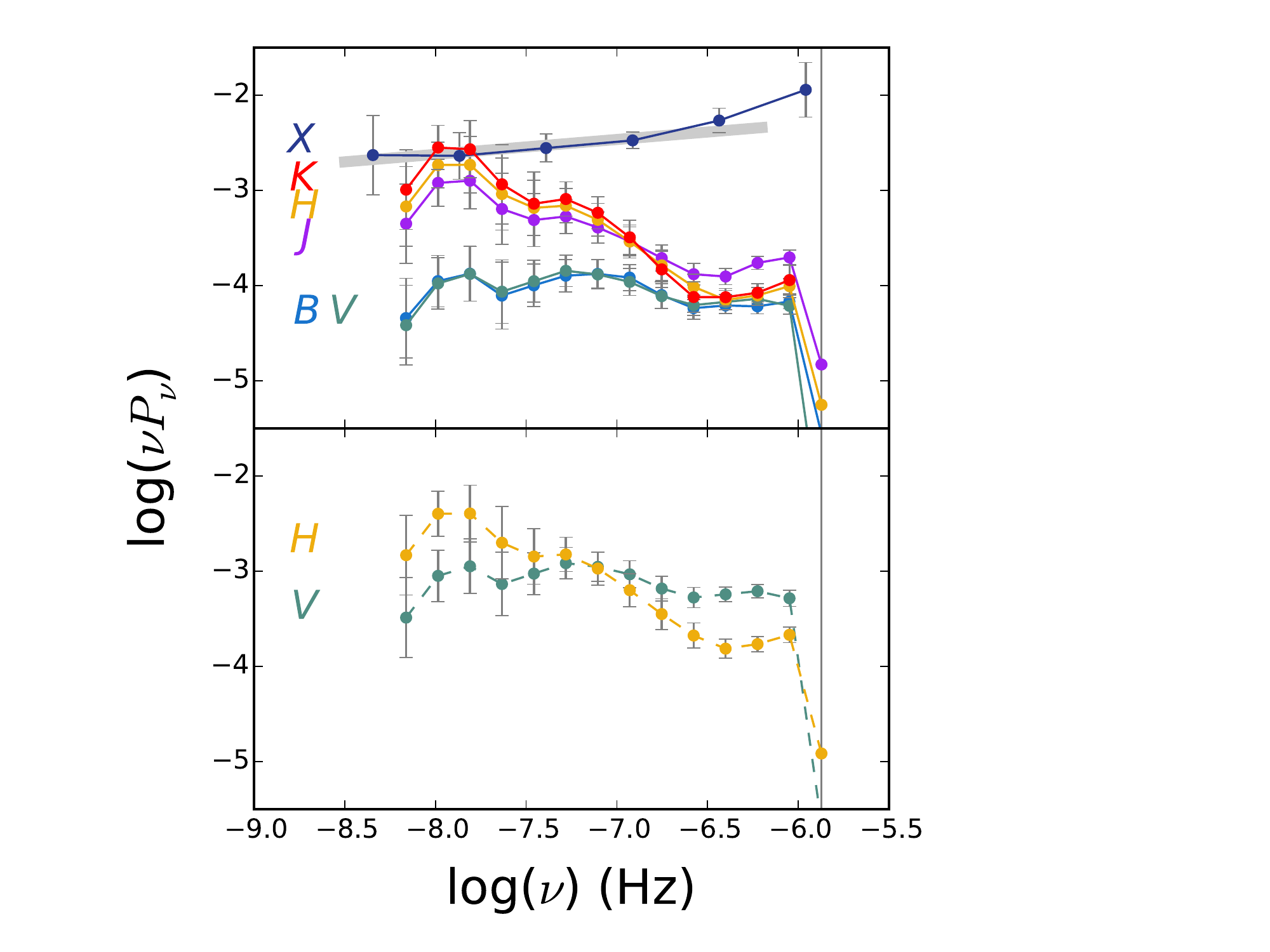}
\caption{Top panel: Power spectral densities (PSDs) obtained from the
  X-ray, optical and near-IR observations for MCG-6-30-15. We also
  include the low-frequency slope previously fitted to the
  RXTE/XMM-Newton X-ray PSD by McHardy et al.~(2005 -- thick gray
  line). Lower panel: PSDs for the V and H bands after correcting for
  host contribution (see text for details).}
\label{mcg63015pds}
\end{figure}

We obtained the power spectral density (PSD) for all bands calculated
using the Mexican hat filter method described in Ar\'evalo et
al.~(2012). The Poisson noise contribution was estimated from the
errors on the fluxes by simulating an error light curve of Gaussian
deviates with zero mean and standard deviation equal to the error in
each flux point. The power spectrum of these error light curves was
calculated with the same method used for the real light curves and was
subsequently subtracted from the total power spectrum. Error bars on
the power spectra represent the expected scatter for different
realizations of a red-noise process. For highly correlated energy
bands (such as B and V bands) this error overestimates the band to
band scatter.
									   
In order to combine the RXTE long time scale and XMM short time scale
observations, we followed McHardy et al.~(2005, 2004), and restrict
the energy range of the XMM light curve to 4-10 keV, whose mean photon
energy match well the 2-10 keV mean energy of a photon from RXTE.

Aliasing from higher frequencies can be important in the X-ray power
spectrum, since its break timescale is about 4 hours (McHardy et
al. 2005) and the observations are on average taken 4 days apart,
i.e., there is significant variability power on timescales much
shorter than those sampled. To estimate the effect of aliasing on the
X-ray power spectrum we simulated X-ray light curves with the power
spectral parameters given in McHardy et al. (2005): a broken power law
($P_{\nu} \propto \nu^{\alpha}$) with slopes $\nu_L = -0.8$ and $\nu_H
= -2.5$ at low and high frequencies, break frequency of $7.6\times
10^{-5}$ Hz, and using a generation bin size of 0.01 days. We
resampled these simulations to match the sampling of the real light
curve and calculated the power spectrum with the same method used for
the real data. The median power spectrum of 100 trials was fitted with
a power-law of free slope and normalization in the range $10^{-8} -
2\times 10^{-6}$ Hz. We varied the input low-frequency slope around
the best fitting value and obtained flatter but comparable slopes in
the resulting power spectra, as expected from the effect of
aliasing. For input slopes of 0.9, 0.85 and 0.8, the measured slopes
were 0.8, 0.76 and 0.69, respectively.  The real X-ray lightcurve gave
a slope of 0.74, which is consistent with an intrinsic slope of 0.85,
similar to the value obtained from a different data set by McHardy et
al. (2005). Aliasing is much less important in the optical and IR
bands, where the sampling rate is sufficient to track the major flux
variations.

Figure \ref{mcg63015pds} presents the PSDs for all bands. Clearly, all
the PSDs present a rather flat distribution in units of $\nu \times
P_{\nu}$, which is recognized as the self-similar flicker-noise region
of the PSD (e.g.~Uttley 2007). The inverted (positive in $\nu \times
P_{\nu}$ units) slope of the X-ray PSD was already seen by McHardy et
al.~(2005), as discussed above. We have included the low-frequency
slope fitted by McHardy et al.~(2005) in Figure \ref{mcg63015pds}.
Our PSD slightly extends the low frequency coverage down to $5 \times
10^{-9}$ Hz. No sign of a secondary break is seen, as expected from
observations of other low black hole mass AGN and black hole X-ray
binaries in the soft state (Done \& Gierlinski 2005; Uttley 2007).

The PSDs of the longer wavelength bands show clearly less power than
that seen in the X-rays. Interestingly, the near-IR PSDs have {\em
  more\/} power than those of the B and V bands, except at frequencies
above $\sim 5 \times 10^{-7}$ Hz where the power becomes comparable.
However, the host galaxy makes a non-negligible contribution to the
optical and near-IR bands, adding to the total flux but not to the
fractional flux variations, therefore decreasing the amplitude in the
observed variability and diminishing the power in the PSD (notice that
extinction, being a multiplicative factor, does not change the
obtained PSD; e.g.~Uttley et al.~2002). We can assess this correction
using the H-band IFU SINFONI data recently presented by Raimundo et
al.~(2013). They find that the total emission within a circular
aperture with diameter 3 arcsec (very close to the aperture used to
determine our light curves) corresponds to $f_{H \star} \sim 3.4
\times 10^{-15}$ erg/s/cm$^2$/\AA\ (in very good agreement with the
mean value of our light curve), of which $\sim 45\%$ corresponds to
the underlying stellar population (Raimundo, private communication).

Extrapolating to the optical region is not straightforward, as we need
to assume a spectral energy distribution for the stellar population.
On the one hand, the S0 nature of the MCG-6-30-15 host might argue for
a rather old population. However, based on the presence of FeII
emission lines in the H-band spectra, Raimundo et al.~(2013) argue for
the presence of a nuclear cluster with a stellar age of $\sim 10^8$
years. For such a young stellar population the contribution to the
V-band will be $\sim 10$ times larger than that observed in the
H-band, i.e., $f_{V \star}^0 \sim 1.5 \times 10^{-14}$ erg/s/cm$^2$/\AA, which
translates onto $f_{V \star} \sim 2.7 \times 10^{-15}$ erg/s/cm$^2$/\AA\ for
an extinction of $E(B\!-\!V)=0.6$ (Reynolds et al.~1995), below the
minimum level seen in our V-band light curve.

The bottom panel in Figure \ref{mcg63015pds} presents the V and H
corrected PSDs after subtracting the host contribution in both
bands. It can be seen that the two PSDs have now a similar power
level, and that they might cross somewhere between $10^{-7.5}$ and
$10^{-7.0}$ Hz, but the differences are within one sigma errors and
the scaling of the PSDs is very sensitive to the host correction just
introduced, which is not well known. Still, it is interesting that
the near-IR bands show such a large amount of variability power when
compared with the optical, since disc variability is expected to
decrease at larger radii.

In some more detail, note that around $10^{-8.0}$ Hz or 2 years the
uncorrected K-band has comparable power to the X-ray band, but
galactic contamination is probably negligible at X-ray energies, so
this power spectrum gives a good estimate of the true fractional X-ray
variability.  The K-band, on the other hand, contains some level of
stellar contamination, and therefore the AGN fractional variability in
this band can only be corrected upwards. Finally, the V-band
variability does not reach these high powers even after correcting for
stellar contamination. The large amplitude of the K-band fluctuations
can be interpreted as reprocessed thermal emission from the
torus. Since this structure does not have and internal source of
heating, it can only respond to the optical, UV and X-Ray continua
emitted by the disc and corona. The long term K-band variations in
MCG-6-30-15 are as large as those of the X-Rays and could also be
responding to the (unobserved) UV, but the optical power seems to be
insufficient to drive the near-IR variability.

\subsection{Correlation Analysis}

\begin{figure*}
\centering
\includegraphics[scale=0.6,angle=0,trim=80 0 0 0]{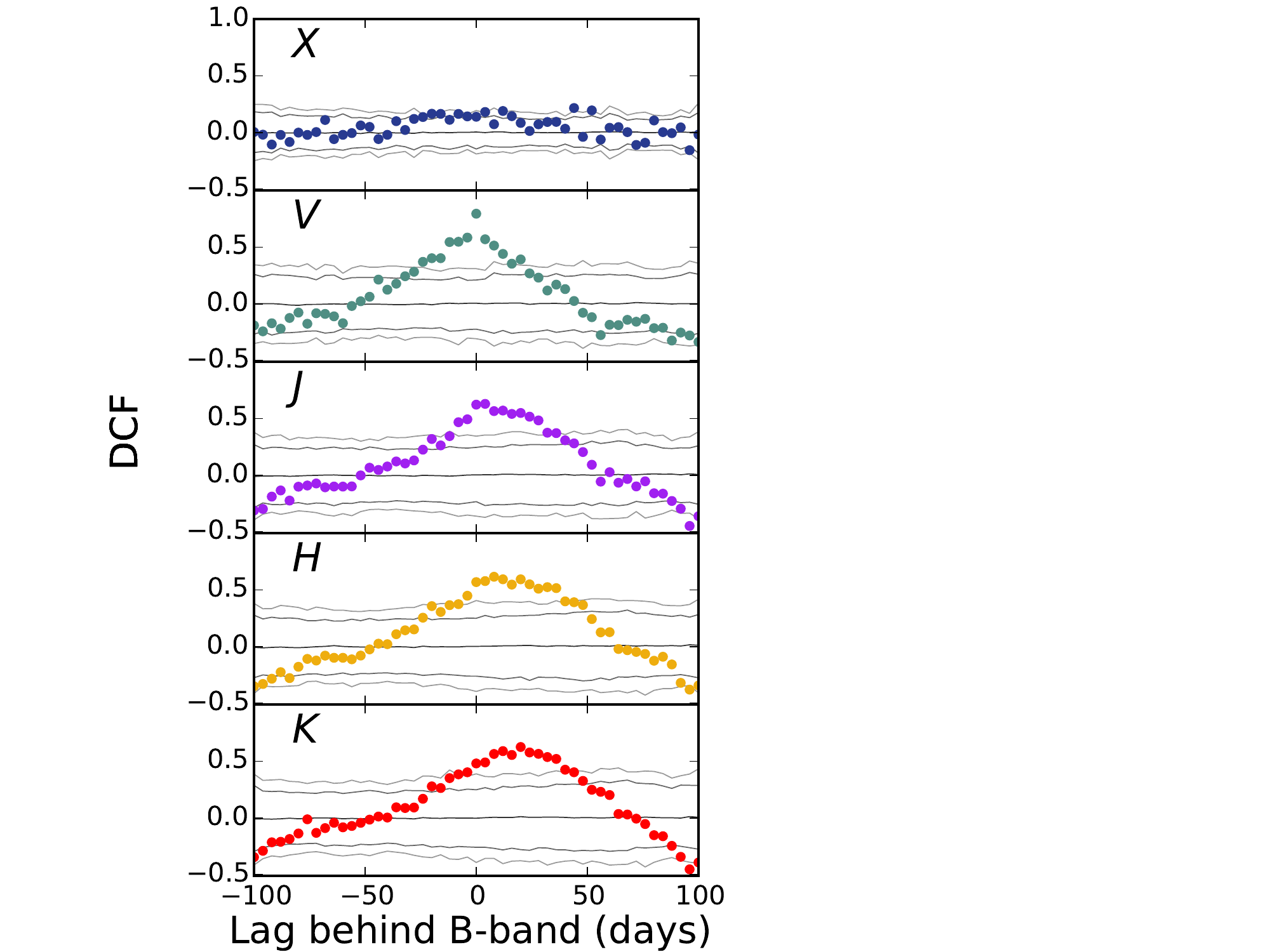}%
\includegraphics[scale=0.6,angle=0,trim=240 0 0 0]{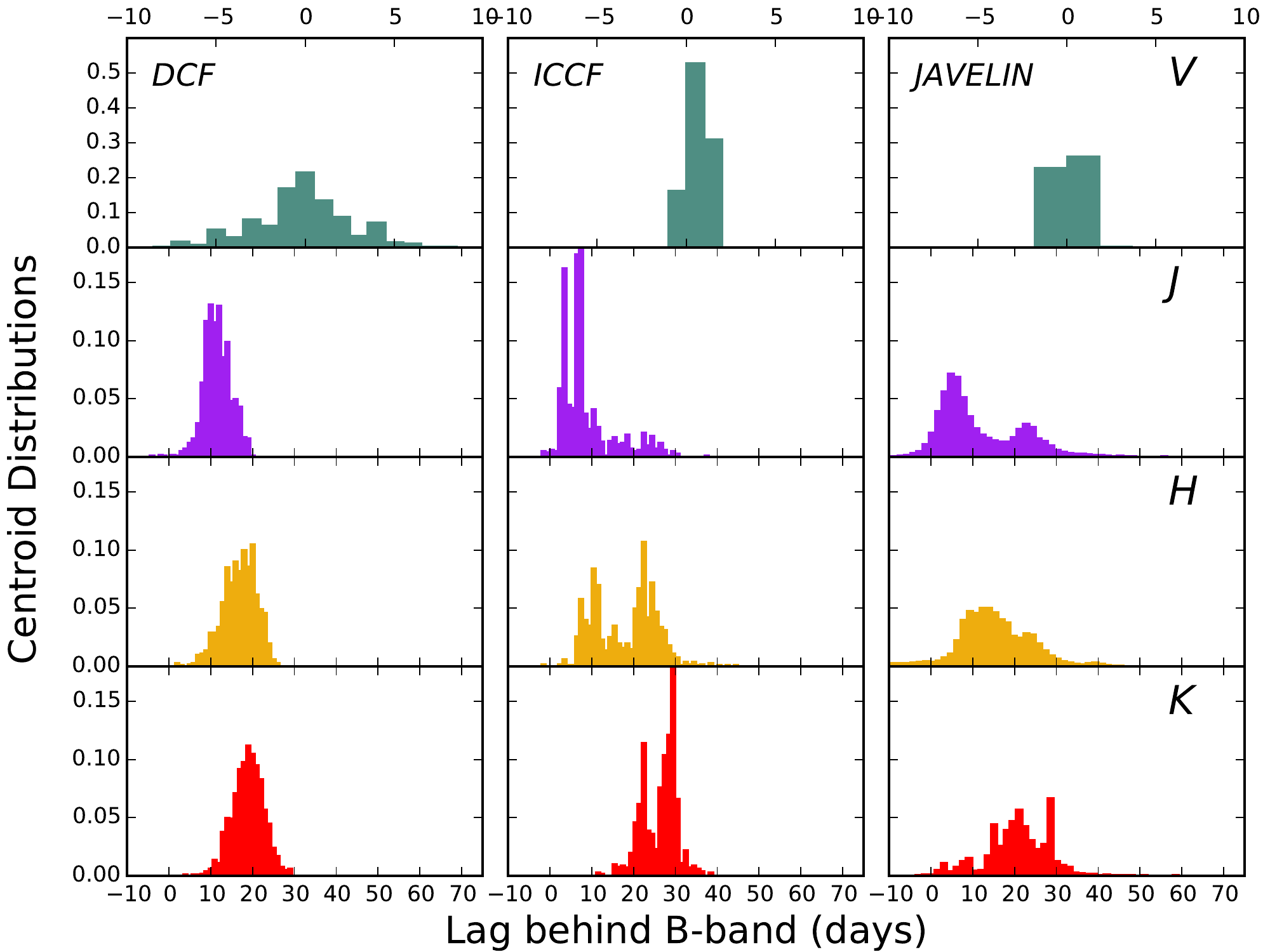}
\caption{{\bf Left:} Correlation coefficient from CDF analysis between
  the B band and all other observed bands for MCG-6-30-15. Positive
  lags mean that the annotated band lags behind B. The continuous gray
  lines represent the mean, 95\% and 99\% upper and lower confidence
  limits. {\bf Right:} DCF, ICCF and JAVELIN centroid distributions
    between the B band and all other optical and NIR observed
    bands. As before, positive lags mean that the annotated band lags
    behind B. Notice the different axis ranges for the top panels.}
\label{mcg63015cc}
\end{figure*}

\begin{table*}
\centering
\caption{Results from the correlation analysis of the full and
  filtered light curves. Lags are measured with respect to the B band
  and are expressed in days. A positive lag means the B band leads the
  variability. JAVELIN $1\sigma$ confidence limits and centroids for
  the discrete correlation functions (DCFs) and interpolated cross
  correlation functions (ICCF) are presented. ICCF centroids are
  determined at a $1\sigma$ level. DCF lag centroids measured at a
  95\%\ confidence level are denoted with a star while other lags are
  determined at a 99\%\ confidence level.}
\begin{tabular}{cccccccc} \hline
  & Javelin Full & ICCF Full        & DCF Full             & DCF $k1$ & DCF $k2$ & DCF $k3$ & DCF $k4$ \\ \hline \smallskip
V & 0.0 :  1.8    & $-0.7 \pm 2.2$ & $-0.4^{+2.1}_{-3.0}$ &$-8.2^{+3.1}_{-2.9}$ & $-5.3^{+4.0}_{-3.9}$      & $-3.8^{+1.9}_{-1.8}$ & $-1.8^{+1.3}_{-0.6}$ \\ \smallskip
J & 4.1 : 25.0    & $13.5 \pm 4.3$ & $11.0^{+3.0}_{-2.9}$ &---               & $10.0^{+3.3}_{-3.3}$       & $9.5^{+2.2}_{-2.6}$  & $10.3^{+2.1}_{-1.9}$ \\ \smallskip
H & 8.6 : 25.1    & $20.0 \pm 4.0$ & $16.7^{+3.7}_{-4.2}$ &---               & $\star 17.6^{+3.9}_{-3.9}$ & $13.6^{+2.8}_{-1.9}$ & $17.0^{+3.3}_{-4.4}$ \\ \smallskip
K & 6.9 : 29.3    & $26.1 \pm 3.7$ & $18.5^{+3.8}_{-3.6}$ &---               & $\star 16.2^{+4.4}_{-4.1}$ & $18.0^{+3.4}_{-3.4}$ & $19.8^{+2.1}_{-2.2}$ \\
\hline
\end{tabular}
\end{table*}

From Figure \ref{mcg63015lcs} it can immediately be seen that a high
degree of flux correlation exists between the optical and the near-IR,
with the short term variability gradually diminishing in significance
towards longer wavelengths. No correction for host galaxy contamination
is introduced. Therefore the overall observed amplitudes of the light
curves should be regarded as a lower limit to the real variations.

We have quantified the degree of correlation between bands using three
methods: the discrete correlation function (DCF) of Edelson \& Krolik
(1988) with confidence limit determinations following Timmer \& Konig
(1995), the interpolated cross correlation function (ICCF) presented
by Peterson et al.~(1998, 2004), and the JAVELIN cross-correlation
method of Zu et al.~(2011, 2013), which models the light curves as a
damped random walk process (DRWP) as prescribed by Kelly et
al.~(2009). While the DCF method does not require any assumption about
the the variability to work, the Timmer \& Konig (1995) technique to
derive its significance requires a previous knowledge of the shape of
the PSD in order to simulate synthetic light curves (see below). On
the other hand, the ICCF is a model-independent estimate of the degree
of correlation. Finally, JAVELIN assumes a particular regime of the
PSD (a DRWP or $P_{\nu} \propto \nu^{\alpha}$ with $\alpha = -2$,
breaking to $\alpha = 0$ at a characteristic frequency) in order to
determine a lag and its significance. No host correction has been
introduced to the light curves during the cross-correlation analysis
using either of the described methods.

We determined the DCFs for the year-long segments of the {\em RXTE\/}
and optical--near-IR light curves and then combined them. In Figure
\ref{mcg63015cc} we present the DCFs obtained between the B-band and
all other observed bands, with 95\% and 99\% confidence limits
following Timmer \& Konig (1995), i.e., by determining the cross
correlation of the observed B-band light curve and a synthetic light
curve of the band of interest simulated according to the power law
shape of its PSD. This process was repeated 1000 times and the DCF
distributions obtained were used to determine the confidence
limits. The lags corresponding to the main peak in the DCF
distributions were estimated using the random sample selection method
of Peterson et al.~(2004), selecting 68 percent of the data points in
the B and near-IR light curves and calculating the DCF centroid for
1000 such trials. The resulting centroid distributions are also shown
in Figure \ref{mcg63015cc} and their mean values and errors are
presented in Table 1. Further details of the procedure can be found in
Ar\'evalo et al~(2008).

\begin{figure}
\centering
\includegraphics[scale=0.45,angle=0,trim=100 0 60 0]{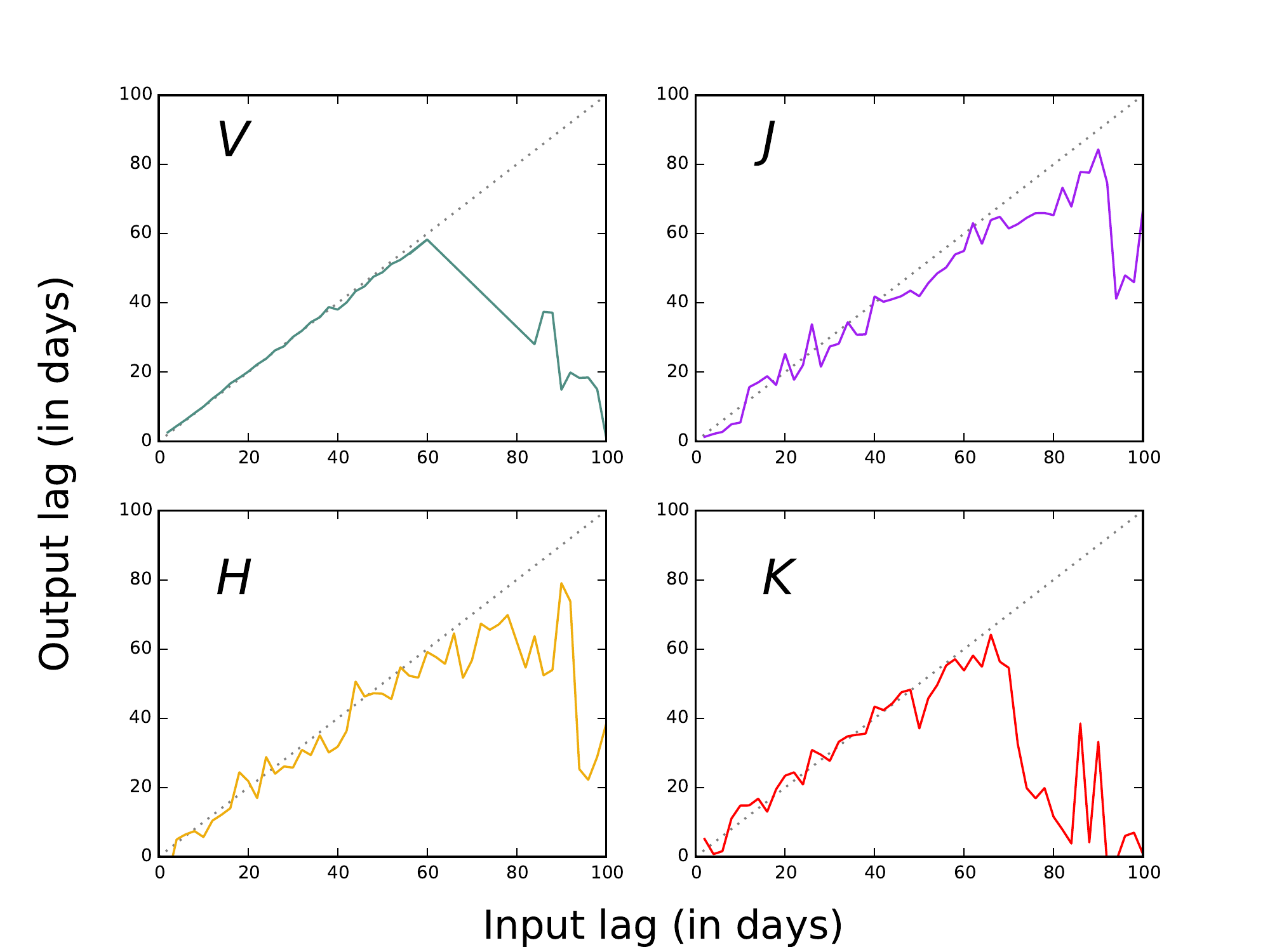}
\includegraphics[scale=0.45,angle=0,trim=60 0 0 0]{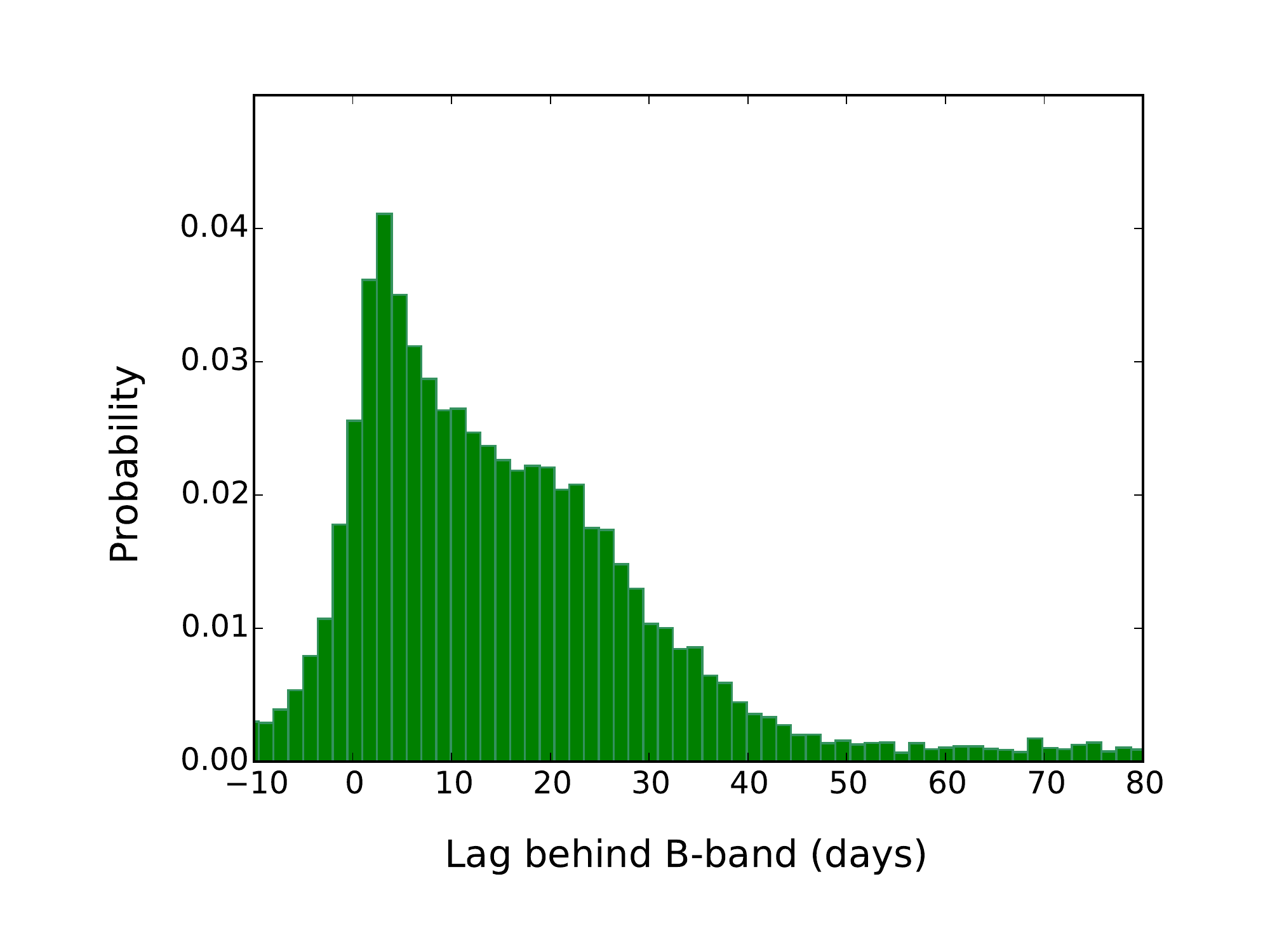}
\caption{Top: JAVELIN results for the cross correlation between the
  observed B-band light curve and the shifted V, J, H and K-bands
  after randomization was applied. See text for details. Bottom:
  JAVELIN lag distribution for simulated B and J light curves
  generated as described in the text.}
\label{javelintest}
\end{figure}

\begin{figure}
\centering
\includegraphics[scale=0.65,angle=0,trim=70 0 20 40]{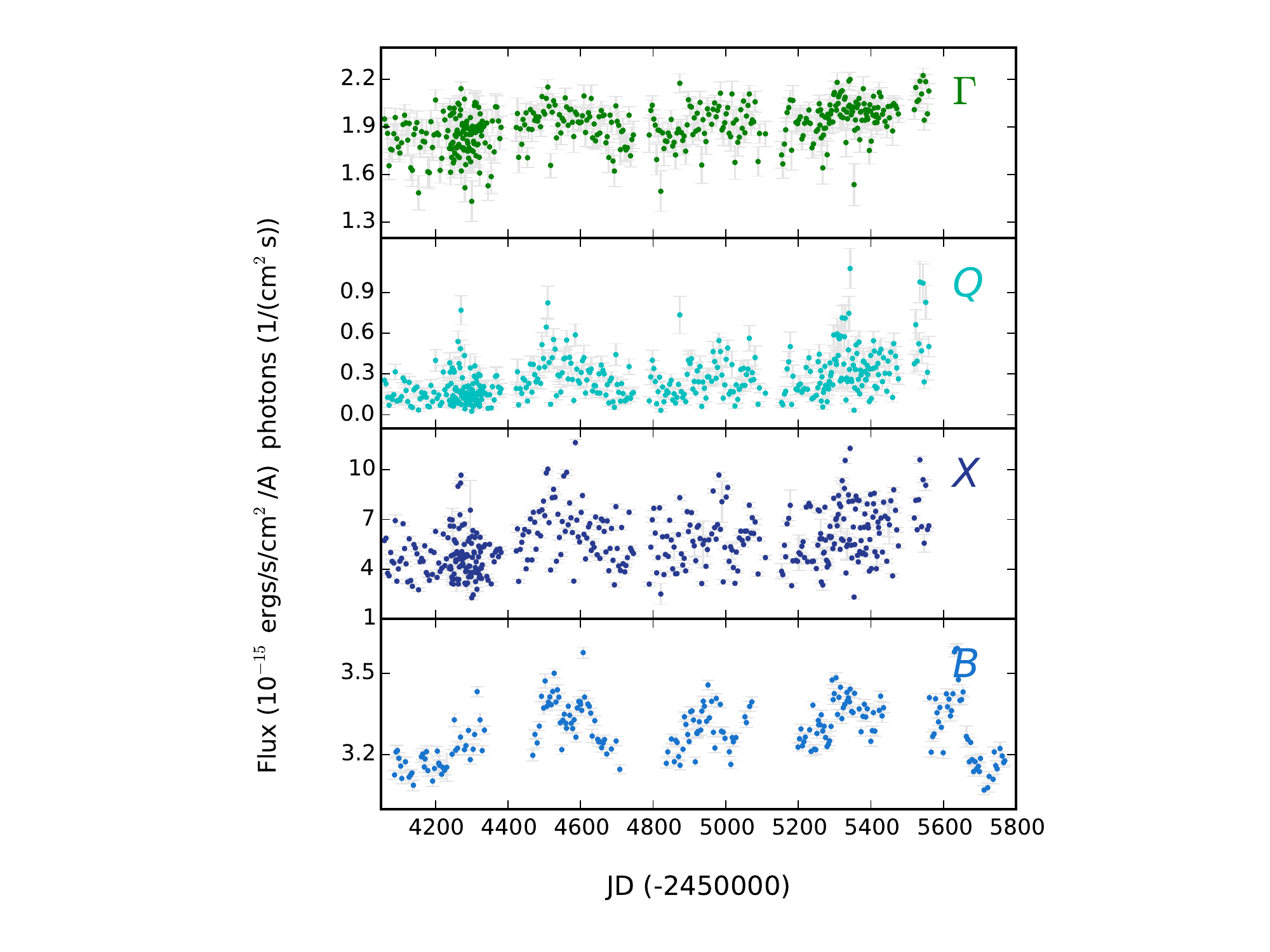}
\caption{$\Gamma$, X-ray photon flux (Q), X-ray flux and B-band flux
  light curves for MCG-6-30-15.}
\label{mcg63015gq}
\end{figure}

ICCF results for the same {\em RXTE\/} and optical--near-IR light
curves were determined using a grid of 1 day for the interpolation of
the data before determining the cross-correlation. For the calculation
of the lag and its uncertainty, interpolated, bootstraped data were
cross-correlated and those results with correlation coefficients
larger than 68\%\ out of 1000 trials where used to find the median lag
and its 1$\sigma$ confidence limits.  The resulting centroid
distributions are presented in Figure \ref{mcg63015cc} and the
measured mean lag are tabulated in Table 1.

For the JAVELIN analysis we followed Shappee et al.~(2014) and
Pancoast et al.~(2014), and calculated the distribution of lags from
10000 Monte Carlo simulations for each separate segment and then
combined the yearly probability distributions. From the resulting
distributions 1$\sigma$ intervals were measured as the lag values that
represented the 1-CL at each end of the distributions, with CL
$=0.6827$. Results are presented in Figure \ref{mcg63015cc} and Table
1.

A quick inspection of Figure \ref{mcg63015cc} confirms that all three
cross correlation methods, DCF, ICCF and JAVELIN, are consistent with
each other, although the DCF centroid distributions show less
structure than the distributions found from the JAVELIN and ICCF
analysis. Interestingly, the J and H bands present some evidence of
double peaks in their distributions, as can be seen in the ICCF and
JAVELIN results. This could be a sign for more than one lag present in
MCG-6-30-15, as will be discussed in more detail in Section 4.2.

Assuming a DRWP as a good description of the observed light curves,
however, might not be appropriate for our study. While DRWP is
characterised by $P_{\nu} \propto \nu^{-2}$, breaking to $P_{\nu}
\propto \nu^{0}$ at lower frequencies, it is clear that our X-ray PSD
is better characterised by $\alpha > -1$ and the optical PSDs are more
consistent with a power distribution given by $\alpha \sim -1$ (see
Figure \ref{mcg63015pds}). The near-IR bands, on the other hand, seem
more consistent with a random walk process. We expect, however, that
the near-IR variability might be the combination of variations coming
from two structures, the disk plus the torus, making the DRWP
assumption also flawed.

To assess this issue we used JAVELIN to determine the lag between the
B-band and the V, J, H and K-bands after these were shifted back to
zero lag using the DCF results reported in Table 1, and then shift
them forwards up to a lag of 100 days using a step of 2 days. For each
calculation, the curve of interest was randomized in its flux assuming
Gaussian distributed errors, and in time, by shifting the date
assuming a flat probability with a width of 10\%\ around the actual
observing date. The results are presented in Figure
\ref{javelintest}. The plots show that the output lags are strongly
under-predicted for delays longer that 50 days, but that they can
recover input lags before that. Since all of our findings report lags
much shorter than this our JAVELIN results should be correct. This is
also supported by the similar behaviour shown by the JAVELIN results
and those from the DCF and ICCF analysis.

In order to investigate whether JAVELIN is indeed able to determine
the presence of more than one lag during the cross-correlation
analysis we performed the following test: B and J-band in-phase
synthetic light curves were computed assuming PSDs with slopes of -1
and -2, similar to the observed values, and break frequencies at 100
and 1000 days, respectively. This ensures that both light curves are
coherent, but with the J-band light curve being much smoother than the
B-band light curve. Next, two versions of the J-band light curves are
computed applying a lag of 5 and 20 days and their average is
determined. The resulting JAVELIN lag distribution between the B-band
and the linearly combined two-lag J-band light curves can be seen in
Figure \ref{javelintest}. Clearly, there is evidence for a double peak
consistent with the lags introduced in the synthetic light curves. To
find what combination of parameters would yield results similar to
those presented in Figure \ref{mcg63015cc} is beyond the scope of this
paper.

From the X-ray and B-band DCF plot in Figure \ref{mcg63015cc} it can
be seen that no significant correlation signal is found. JAVELIN
analysis did not give a clear result (plot is not shown) which could
be related to the problem of adopting a DRWP as a description of our
X-ray observations.

The DCF result is not totally unexpected as the X-rays show very fast
variability and our {\em RXTE\/} and optical--near-IR light curves are
not suitable to determine a correlation between these bands. A
correlation was previously determined between the X-ray emission and
the 3000--4000 \AA\ U-band of the Optical Monitor on-board XMM-Newton
(Ar\'evalo et al., 2005). The observations corresponded to snapshots
800 seconds long separated by gaps of 320 seconds. Such fast
monitoring allowed us to determine a lag of $1.85^{+0.52}_{-0.75}$
days, with the U-band leading the X-rays. In that case the X-ray light
curve tracked lower energies since the bulk of the photons recorded by
the XMM pn camera are below the 2 keV threshold of RXTE and are
therefore dominated by the soft excess present in MCG-6-30-15. In
fact, the soft excess has been shown to correlate better with the
optical bands than the hard X-rays in at least one source with good
quality UV, soft and hard X-ray light-curves (Mrk509; Mehdipour et
al.~2011). In fact, a re-analysis of the XMM-Newton data by Smith \&
Vaughan (2007) did not report a significant lag. Smith \& Vaughan
(2007) argue that the differences with Ar\'evalo et al.~(2005) are due
to the different methods used to extract the light curves and a more
conservative treatment of the correlation analysis.

We also obtained the cross correlation between the B-band and the
X-ray photon flux light curve, Q, which is presented in Figure
\ref{mcg63015gq}. In the framework where the corona is cooled by
Comptonization of (the unseen) UV and optical photons, significant
changes in the UV/optical flux should be correlated with changes in
the X-ray flux. However, the corona energetics also respond to changes
in the UV/optical flux, with higher fluxes inducing more efficient
electron cooling. This in turn will mean that B-band photons will gain
less energy from the corona and therefore the shape of the X-ray
spectrum will become steeper. This trend is seen in Figure
\ref{mcg63015gq}, where the photon index $\Gamma$ of the 3-10 keV
energy range (i.e., for flux $\propto \nu^{-\Gamma}$ photons/s/cm$^2$)
grows steadily during the monitoring from a value of $\sim 1.8$ at the
start of the campaign to $\sim 2.0$ towards the end.

\begin{figure}
\centering
\includegraphics[scale=0.48,angle=0,trim=45 0 0 130]{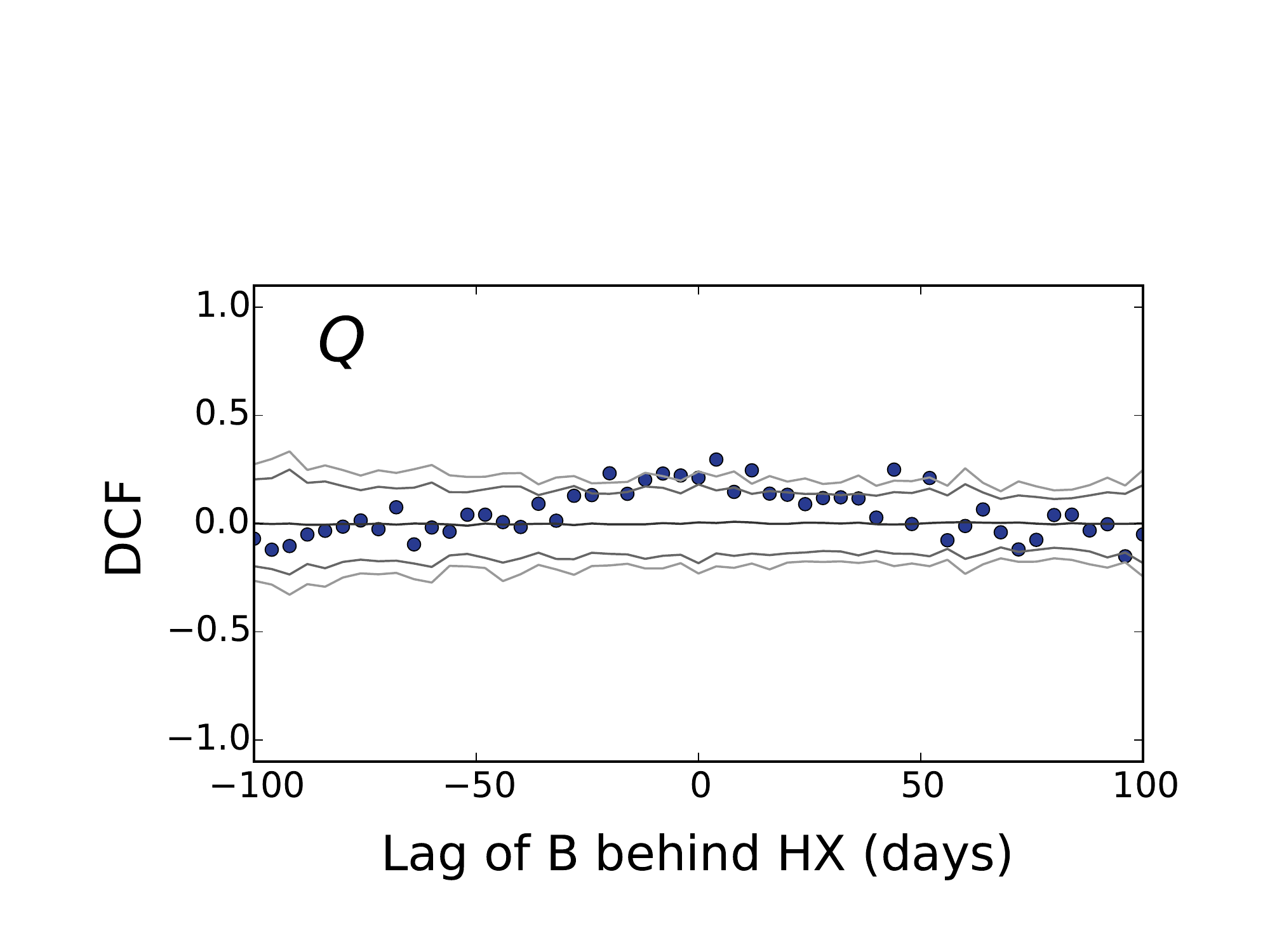}\\
\includegraphics[scale=0.48,angle=0,trim=45 10 0 0]{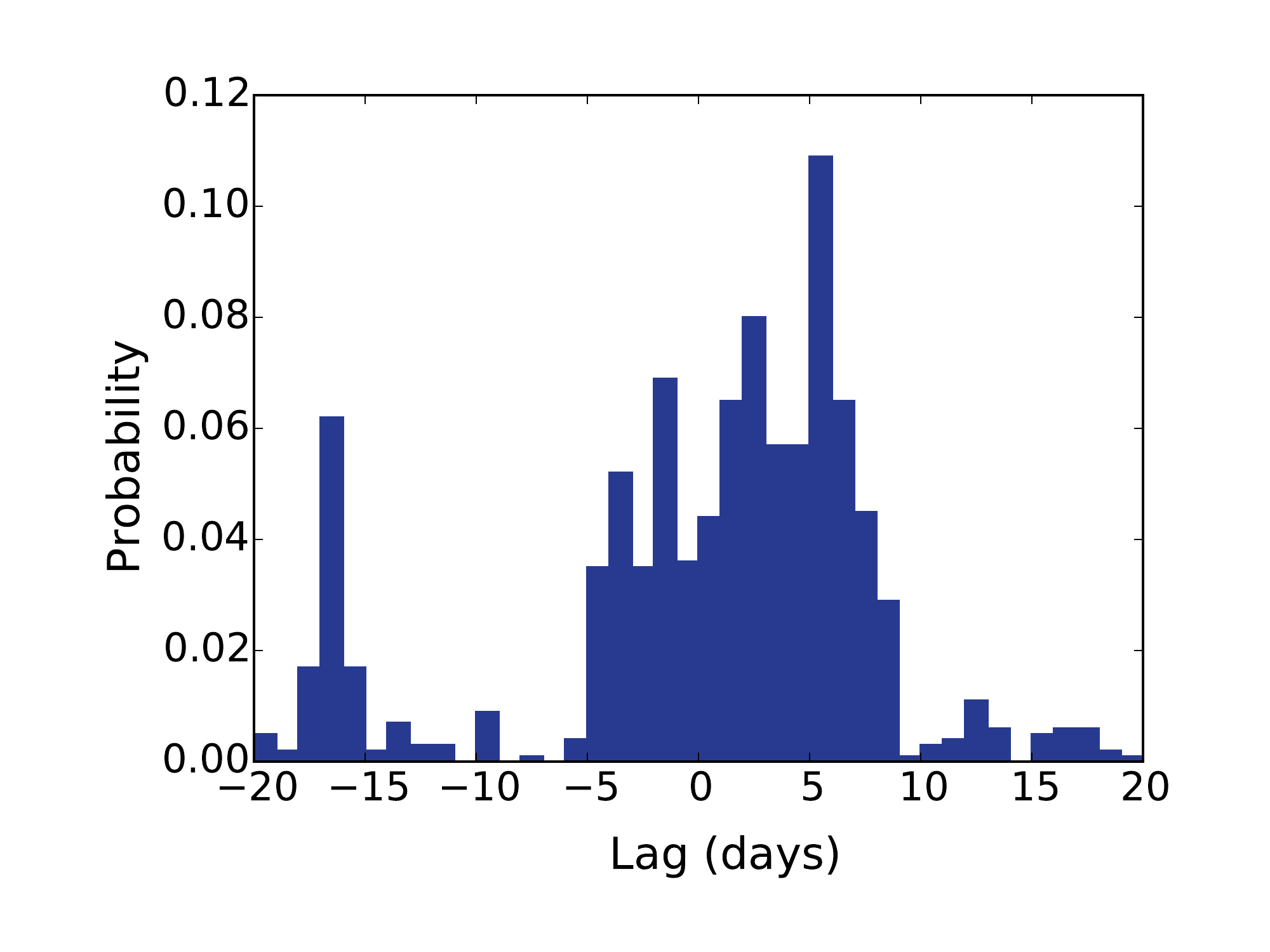}
\caption{{\bf Top:} Cross corelation between the Q and B-band from the
  DCF. {\bf Bottom:} Cross corelation between the Q and B-band from
  the ICCF. }
\label{mcg63015dcfq}
\end{figure}

As shown by Nandra et al.~(2000), in cases of a variable $\Gamma$ it
is expected that the photon flux Q would be more closely correlated
with the UV or optical flux, because of the one-to-one nature of the
Compton scattering between photons and electrons. In Figure
\ref{mcg63015dcfq} we present the cross correlation between Q and the
B-band obtained using the DCF and the ICCF methods. As can be seen,
there is good evidence for a correlation with a lag around
zero. Unfortunately, the centroids are found with very large spreads
of $8^{+6}_{-48}$ and $2.1 \pm 20.1$ days for the DCF and ICF methods,
respectively, with the Q-band leading. The correlation, however, is
present, with a 100\%\ of the (1000) centroid calculations performed
during the ICCF trials being successful and a mean peak correlation
coefficient of $0.48 \pm 0.04$.

On the other hand, there is a clear correlation signal between the
near-IR bands and the optical bands. The rather flat near-IR DCFs in
Figure \ref{mcg63015cc} resemble those of NGC~3783, where a very broad
and flat correlation was interpreted as resulting from the sum of two
variable components varying on different time scales: a rapidly
varying disc and slower dusty torus (Lira et al., 2011). 

\begin{figure}
\centering
\includegraphics[scale=0.45,angle=0,trim=0 0 0 0]{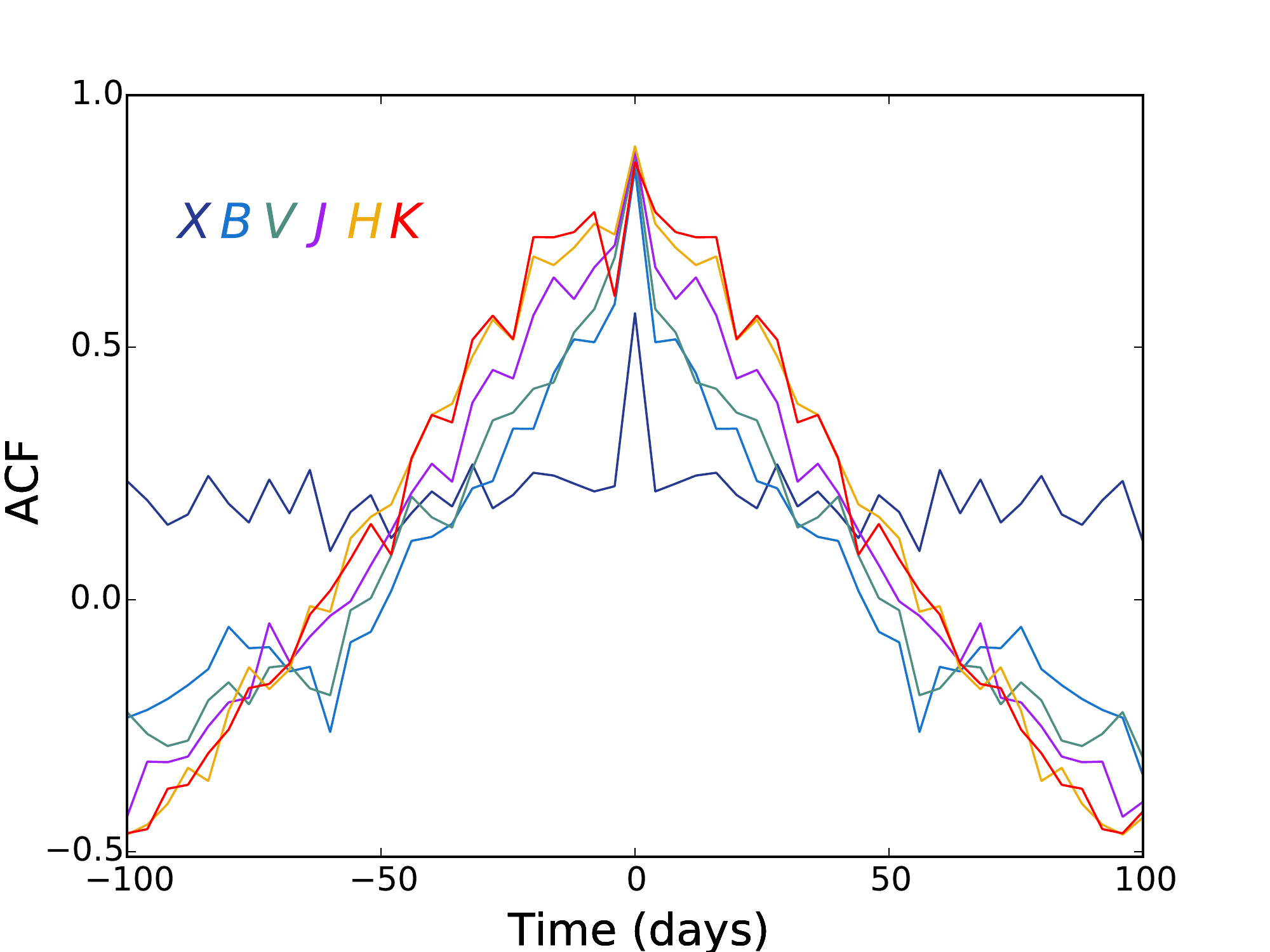}%
\caption{Auto correlation functions (ACFs) of all the light curves
  presented in Figure \ref{mcg63015lcs}. From the inner-most to the
  outer-most curves, the ACFs correspond to the X, B, V, J, H and
  K-bands.}
\label{mcg63015acf}
\end{figure}

We can test the presence of more than one component by comparing the
width of the auto correlation functions (ACFs) in different bands. If
one band drives the variability in the remaining bands, then it is
expected that its ACF should be the narrowest, with the remaining band
ACFs been broadened by the response of the system. Long time responses
will correspond to broad ACFs, and most likely, to more extended
emitting regions. The X-ray, B, V, J, H and K-band ACFs are presented
in Figure \ref{mcg63015acf}. As can be seen, there is a systematic
broadening of the ACFs when going from the shortest to the longest
wavelengths, with the X-ray ACF been particularly sharp and narrow. At
the same time, the ACFs of the B and V bands are very similar, while
the H and K ACFs also look very much alike and are clearly the
broadest of all. This might indicate the presence of a large
reprocessor, like the dusty torus, significantly contributing to the
variability in the H and K bands.

The ICCF and JAVELIN results for the J, H and possibly the K-band are
complex and hint at the presence of more than one distinct variable
component. In fact, the centroid values presented in Table 1 should be
considered as a poor representation of the multiple peaks observed.

\subsection{Filtered Light Curves}

\begin{figure}
\centering
\includegraphics[scale=0.4,angle=0,trim=50 0 0 60]{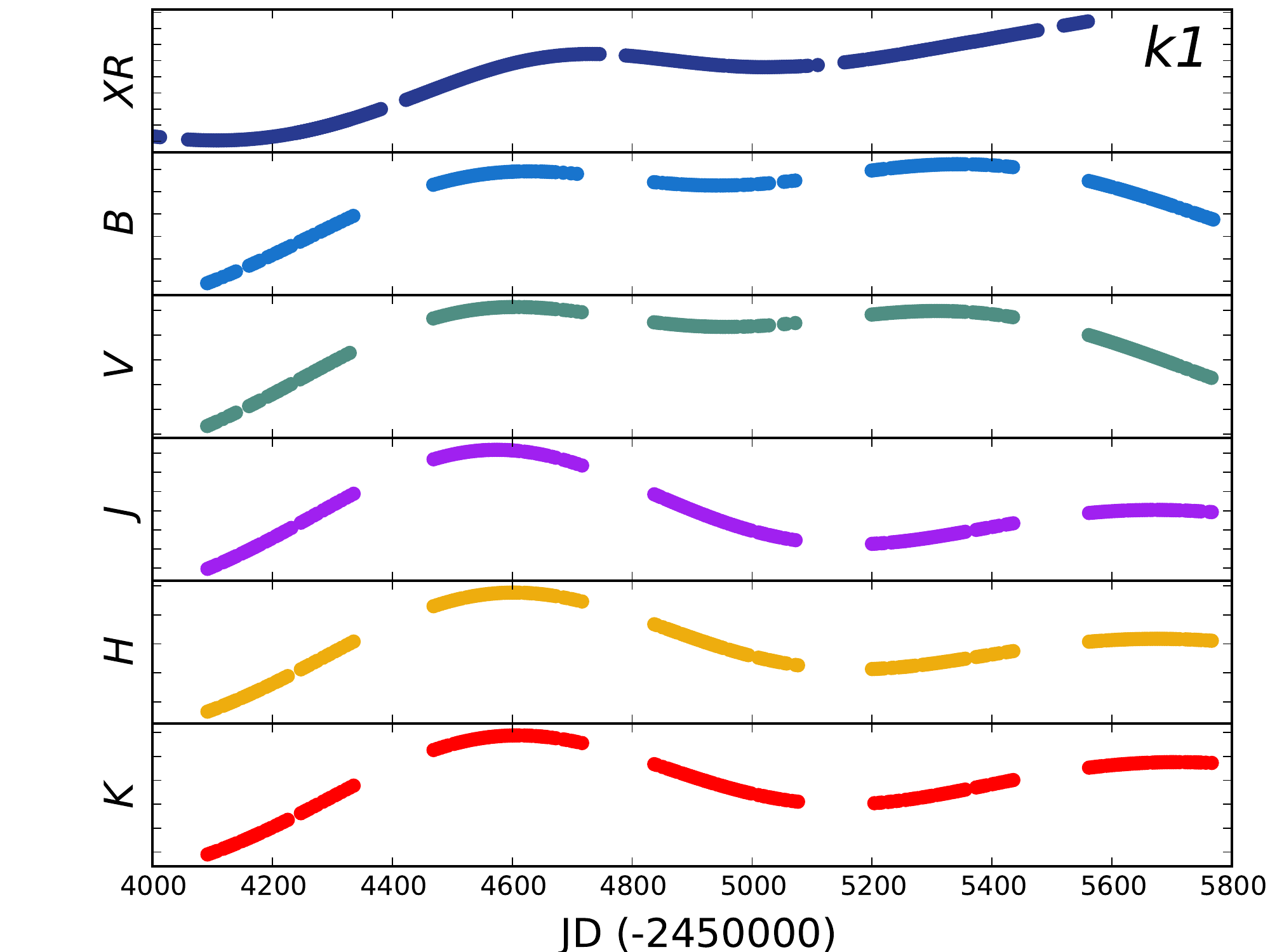}\\
\includegraphics[scale=0.4,angle=0,trim=50 0 0 40]{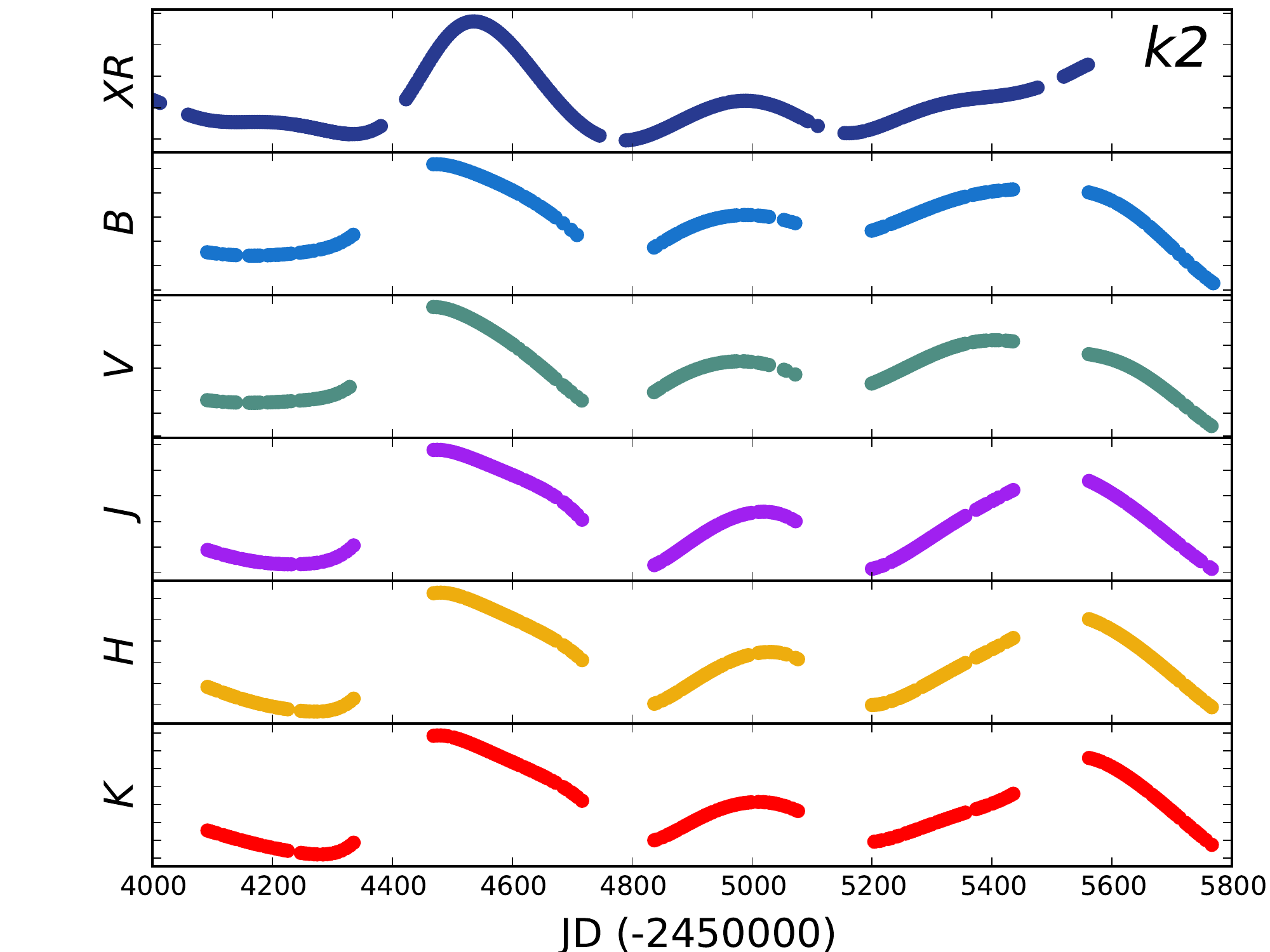}\\
\includegraphics[scale=0.4,angle=0,trim=50 0 0 40]{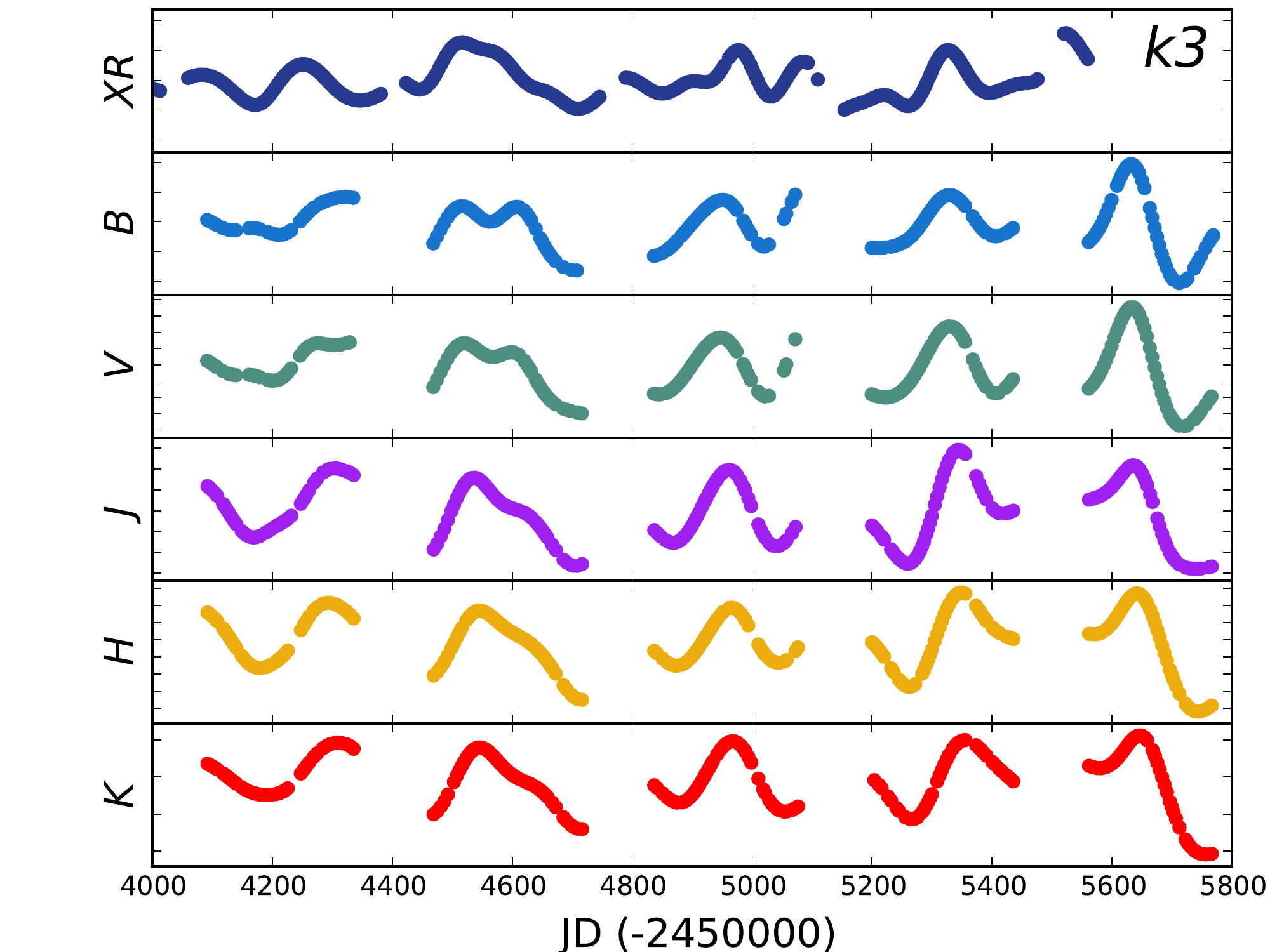}\\
\includegraphics[scale=0.4,angle=0,trim=50 0 0 40]{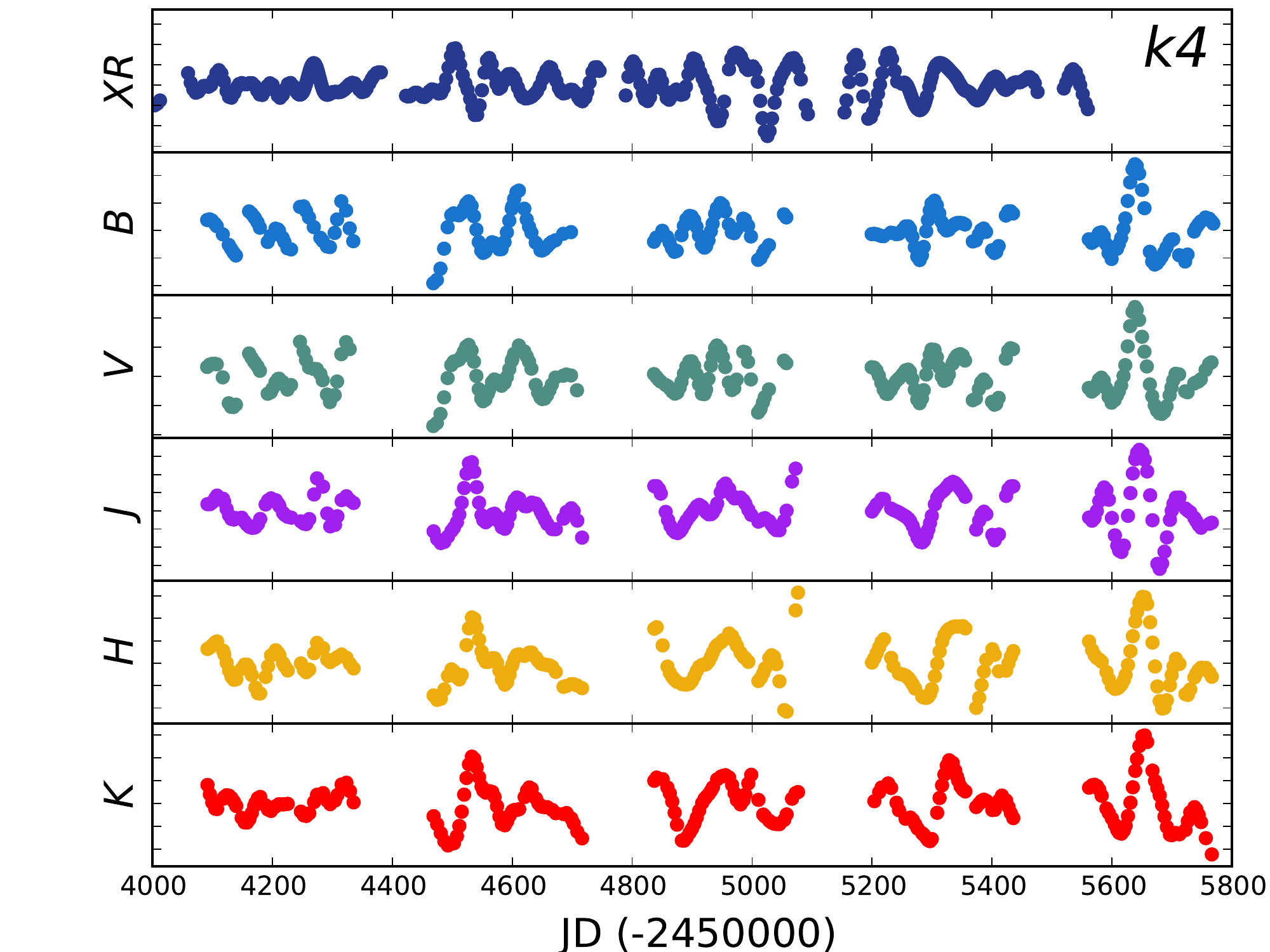}
\caption{Filtered light curves for the X-ray, B, V, J, H, and K bands,
  and for frequency ranges $k1=7\! \times\! 10^{-9}\!-2\! \times\!
  10^{-8}\!$ Hz, $k2=2\! \times\! 10^{-8}\!-6\! \times\! 10^{-8}\!$
  Hz, $k3=6\!  \times\!  10^{-9}\!-2\! \times\! 10^{-7}\!$ Hz, and
  $k4=2\! \times\! 10^{-7}\!-6\! \times\! 10^{-8}\!$ Hz.}
\label{mcg63015k3lcs}
\end{figure}

We filtered the optical and near-IR light curves using the method
described by Ar\'evalo et al.\ (2012). In short, the method consists
of filtering the data using a 'mexican-hat' filter which suppresses
fluctuations with time scales much larger or much smaller than a
characteristic time scale. The method is also able to deal with gaps
in the data, which are masked out during the analysis. In our case we
did not mask out any data but instead worked on year-long segments.

Four light curves were produced with characteristic frequency
intervals of $k1=7\! \times\! 10^{-9}\!-2\! \times\! 10^{-8}\!$ Hz
(560-1680 days), $k2=2\! \times\! 10^{-8}\!-6\! \times\! 10^{-8}\!$ Hz
(186-560 days), $k3=6\! \times\!  10^{-9}\!-2\! \times\! 10^{-7}\!$ Hz
(62-186 days), and $k4=2\! \times\! 10^{-7}\!-6\! \times\! 10^{-8}\!$
Hz (21-62 days). These intervals are equally spaced in log frequency
and the light curves were determined applying the 'mexican-hat' filter
to the PSDs. Figure \ref{mcg63015k3lcs} shows the light curves
obtained for all our bands for the four frequency ranges.

We determined the cross correlation of the filtered light curves using
the DCF method. Unfortunately, it is not possible to apply the JAVELIN
code in this case as the random walk modelling of the light curves is
only valid when describing the full data. As before, no significant
cross correlation results were obtained for the X-ray filtered curves
and the optical or near-IR filtered curves.

\begin{figure}
\centering
\includegraphics[scale=0.65,angle=0,trim=40 0 0 0]{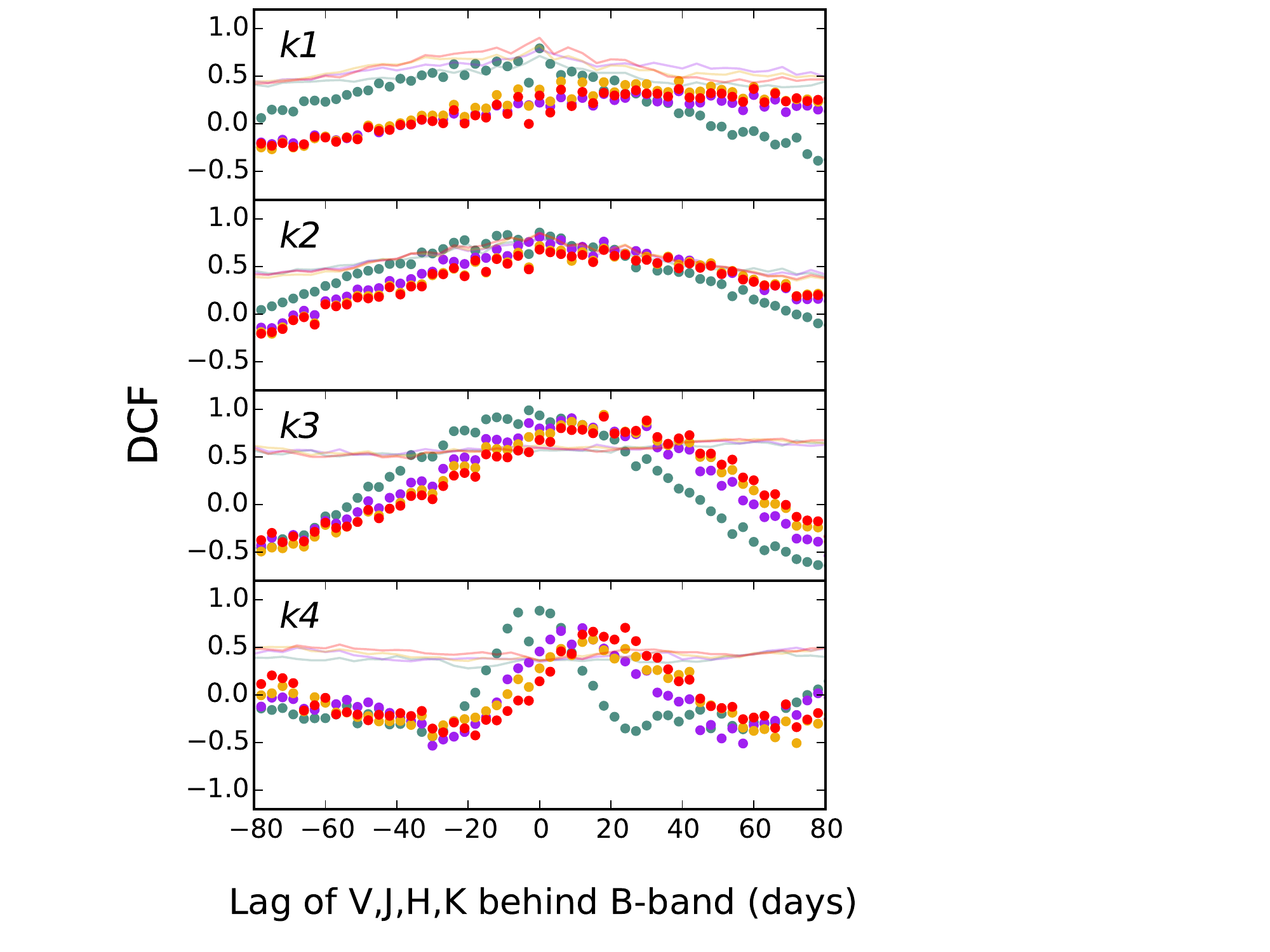}
\caption{Correlation coefficients from DCF analysis between the B and
  VJHK bands for MCG-6-30-15 from the lowest (top) to the highest
  (bottom) frequency ranges. The solid lines represent the
  99\%\ colour-coded upper confidence limits for the different DCFs.}
\label{mcg63015kdcfs}
\end{figure}

Figure \ref{mcg63015kdcfs} shows the DCFs between the B and V, J, H
and K bands corresponding to all the frequency ranges. Only upper
99\%\ confidence limits are included in the plots this time. It can be
seen that the level of significance of the DCFs is higher at higher
frequencies. In particular the first frequency range, corresponding to
$\sim 3$ years, which is hardly sampled by our monitoring programme,
does not give significant lags for any of the near-IR bands.  However,
a significant lag can be measured for the V-band because of its
extreme similar behaviour to the B-band. For the $k2$ frequency range
the V and J bands show DCFs significant at the 99\%\ level, while for
$k3$ and $k4$ all DCFs are significant at this level. At a
95\%\ significance level, all DCFs are significant for frequency
ranges $k2$, $k3$ and $k4$. Centroids are presented in Table 1.

A quick examination of Table 1 shows that for a given frequency range
the near-IR bands respond with characteristicly longer lags when going
towards longer wavelengths (i.e., when reading the table
vertically). For instance, the lag nearly doubles when going from the
J to the K band for the $k3$ and $k4$ frequency ranges. At the same
time there is no clear trend in the different lags for the same band
measured at different frequency ranges (i.e., reading the table
horizontally). So, for example, the J band has a lag consistent with
$\sim 10$ days for all frequency ranges, while the K band hints at
lags that increase monotonically from $k2$ to $k4$. This implies that
the reprocessor is responding equally at all time scales, and the lags
only depend on the observed band.

Because of the similar variability pattern observed in the B and V
bands, the cross-correlations of their frequency filtered light curves
show larger correlation coefficients and smaller lag errors than those
obtained in the near-IR. It is possible, then, to look at some trends
that the lags exhibit with frequency. First of all, it can be noticed
that all the lags determined using the DCF method are negative and
that the values become more negative towards lower frequencies. This
is likely due to contamination of emission from the Broad Line Region
(BLR) to our optical photometry. The strongest line in the wavelength
range covered by our observations ($\sim 3900-5900$ \AA) is H$\beta
\lambda4861$, which falls in B-band filter. This would explain why the
lag gets shorter with increasing frequency range, as the component
from the slowly varying BLR will be more pronounced in the $k1$ than
in the $k4$ light curve. Notice also that the full light curve shows
the shortest lag. This is likely because the frequency ranges examined
leave out a large fraction of the variability power seen in the
optical, which concentrates at shorter frequencies than those found in
the $k4$ range.

Since there is very little line contamination in the V-band we
conclude that the B-band light curve is delayed. However, given the
small negative lag measured from the cross correlation of the full
light curves ($-0.4^{+2.1}_{-3.0}$ and $-0.7 \pm 2.2$ days for the
DCF and ICCF analysis, respectively, see Table 1), we will not attempt
to correct for this effect.

\section{Discussion}

\subsection{The nature of the X-ray variability}

The presence of the broad Fe K$\alpha$ line in the X-ray spectrum of
MCG-6-30-15 has made this object one of the most intensively studied
AGNs in the sky. Modelling of this feature as a gravitationally
redshifted line scattered off hot, optically thick material located at
a distance below $10 R_g$ from the central black hole provides some of
the best evidence for the presence of an accretion disc and opens the
possibility of studying the behaviour of matter and radiation in a
strong gravity environment (e.g., Tanaka et al.~1995; Iwasawa et
al.~1996; Guainazzi et al.~1999; Lee et al.~1999; Vaugh \& Edelson
2001; Wilms et al.~2001; Fabian et al.~2002; Shih et al.~2002; Fabian
\& Vaughan 2003; Marsumoto et al.~2003; Vaughan \& Fabian 2004;
Miniutti et al.~2007; Kara et al.~2014).

However, some authors have also proposed alternative models where the
line can be explained by the presence of a continuum affected by
complex absorption by ionized material, or a 'warm absorber' (Inoue \&
Matsumoto 2003; Miller, Turner \& Reeves 2008, 2009). Crucially for
our study, using a warm absorber composed of several zones with
varying partial covering of the central source, Miller, Turner \&
Reeves (2009) and Miyakawa et al.~(2009) propose to explain not only
the X-ray spectrum of MCG-6-30-15, but also most of its
variability\footnote{It is important to notice, however, that all
  these studies have focused on explaining the spectral changes in the
  X-ray observations, with none of them attempting to also reproduce
  the observed light curves. In fact, a complex warm absorber present
  in MCG-6-30-15 is known to show variability with time scales of a
  few hours (Reynolds et al.~1995), but it is thought to be mostly
  confined to energies below 3 keV, with some absorption features seen
  around the Fe K$\alpha$ line (Chiang \& Fabian 2011), so this
  component is of no interest in the present analysis.}.

However, there is one piece of evidence that argues against the above
scenario and that we can explore here: the rms-flux relation of the
X-ray emission. Uttley \& McHardy (2001) and Uttley et al.~(2005) have
demonstrated the non-linear nature of the X-ray variability in AGN. In
fact, they show that variability is a multiplicative process, while
additive processes, such as the combination of independent occultation
episodes, can be ruled out. Empirically, this translates into the
'rms-flux' relation. In other words, the multiplicative nature of the
variability predicts a linear correlation between the flux level and
its standard deviation. 

Using a long look observation of MCG-6-30-15, Vaughan et al.~(2003),
already presented a clear 'rms-flux' linear correlation for this
source. We have combined our RXTE observations with those presented in
McHardy et al.~(2005) post year 2000, and new XMM observations
(Marinucci et al.~2014; Kara et al.~2014) with those previously
presented in Vaughan et al.~(2003) to determine two 'rms-flux' plots,
following the method of Vaughan et al.~(2003). Since the temporal
resolution of the RXTE and the XMM observations are very different,
each plot presents the results for each space-craft.

\begin{figure}
\centering
\includegraphics[scale=0.45,angle=0,trim=30 0 0 0]{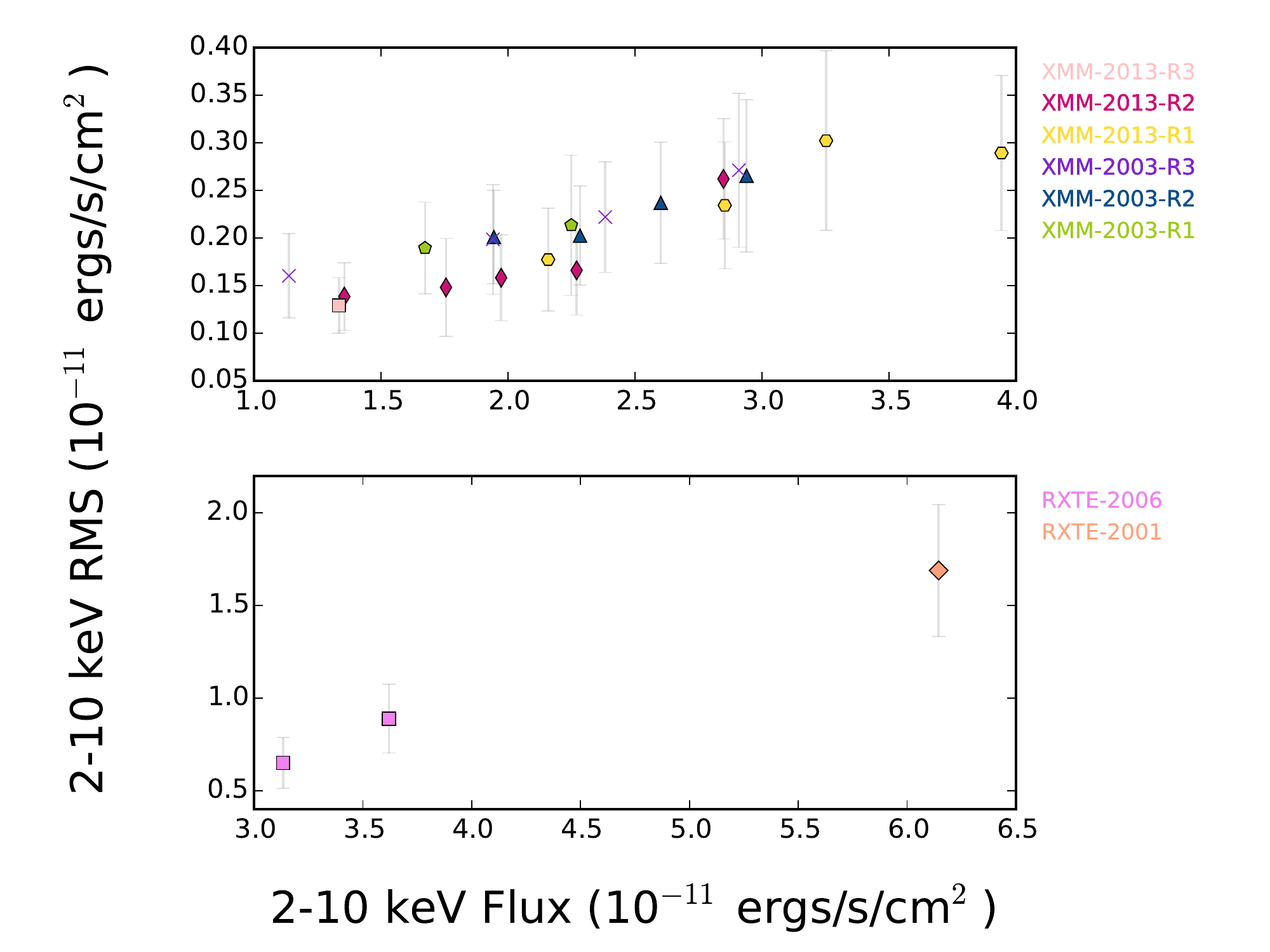}
\caption{RMS-flux correlations for MCG-6-3015. {\bf Top:} XMM-Newton
  observations obtained in 2001 and 2013. Each observation was divided
  into their individual XMM 'revolutions'. {\bf Bottom:} RXTE
  observations obtained by McHardy et al.~(2005) and this work.}
\label{rms_flux}
\end{figure}

The XMM observations were binned into 100 second intervals, while the
observing cadence of the RXTE light curves (approximately 2 and 4 days
for the McHardy and our data, respectively) was not changed. The
weighted mean flux and its standard deviation were determined from
consecutive groupings of 15 bins each. To reduce scatter, further
binning of 15 such flux-rms pairs was obtained. The results are
presented in Figure \ref{rms_flux} where the vertical ``error'' bars
represent the scatter around a given rms value (i.e., they do not
correspond to the error of the mean rms). The linear correlation is
clear and we confirm that multiplicative nature of the X-ray
variability in MCG-6-30-15 in both temporal regimes.

\subsection{Disc and torus variability at optical and near-IR wavelengths}

The discrete correlation function between the B and the near-IR bands
is seen in the left hand side of Figure \ref{mcg63015cc}. The DCFs
show broad, flat-top peaks, which might suggest more than one variable
component contributing to the signal. This has already been seen in
NGC~3783 and interpreted as the disc and the dusty torus
simultaneously contributing to the near-IR DCFs, with the disc
dominating at smaller lags and the torus dominating at longer lags
(Lira et al.~2011). The ICCF and JAVELIN results also suggest the
presence of more than one component in the J and H-band lag
probability distribution seen in the right hand side of Figure
\ref{mcg63015cc}, and even perhaps in the K-band.

In the case of NGC~3783 the presence of the torus is also corroborated
by a clear near-IR hump in its spectral energy distribution (Lira et
al.~2011). Unfortunately, the very high extinction towards the nucleus
of MCG-6-30-15 implies that the determination of its optical to
near-IR spectral energy distribution is rather unfeasible. However,
the H and K bands can give a good idea of the intrinsic spectral shape
since they are less susceptible to extinction, and the stellar
populations show a particularly homogeneous flux ratio between these
two bands ($f_H/f_K \sim 2.5$), regardless of the stellar
age. Adopting a stellar contribution of 45\%\ in the H (see Section
3.1), a 18\%\ K band contribution is then found. For $E(B-V)=0.6$, we
find that absorption and host corrected H and K fluxes are $\sim 3.1$
and $3.5 \times 10^{-15}$ ergs/s/cm$^2$/\AA, respectively, hinting at
a red slope and therefore suggesting the presence of hot dust in the
MCG-6-30-15 nuclear region.

We will attempt to isolate the emission from the disc from that from
the torus by using the distinct 'disc' peaks observed in the JAVELIN
probability distributions, this is, the V peak and the first peaks
seen in the J and H distributions. We will assume that the K-band
JAVELIN main peak corresponds to a 'torus' lag. By fitting a Gaussian
to these features we were able to determine their characteristic
centre and dispersion. The values are: V-disc-lag = $-0.8 \pm 0.9$,
J-disc-lag = $4.9 \pm 3.8$ H-disc-lag = $11.4 \pm 4.6$, J-torus-lag =
$21.9 \pm 4.2$, H-torus-lag = $22.1 \pm 4.5$ and K-torus-lag = $19.6
\pm 4.9$. These values suggests that the outer disc is truncated at
distances $\ga 15$ light-days, with the torus appearing at $\la 20$
light-days.

\begin{figure}
\centering
\includegraphics[scale=0.45,angle=0,trim=50 0 0 0]{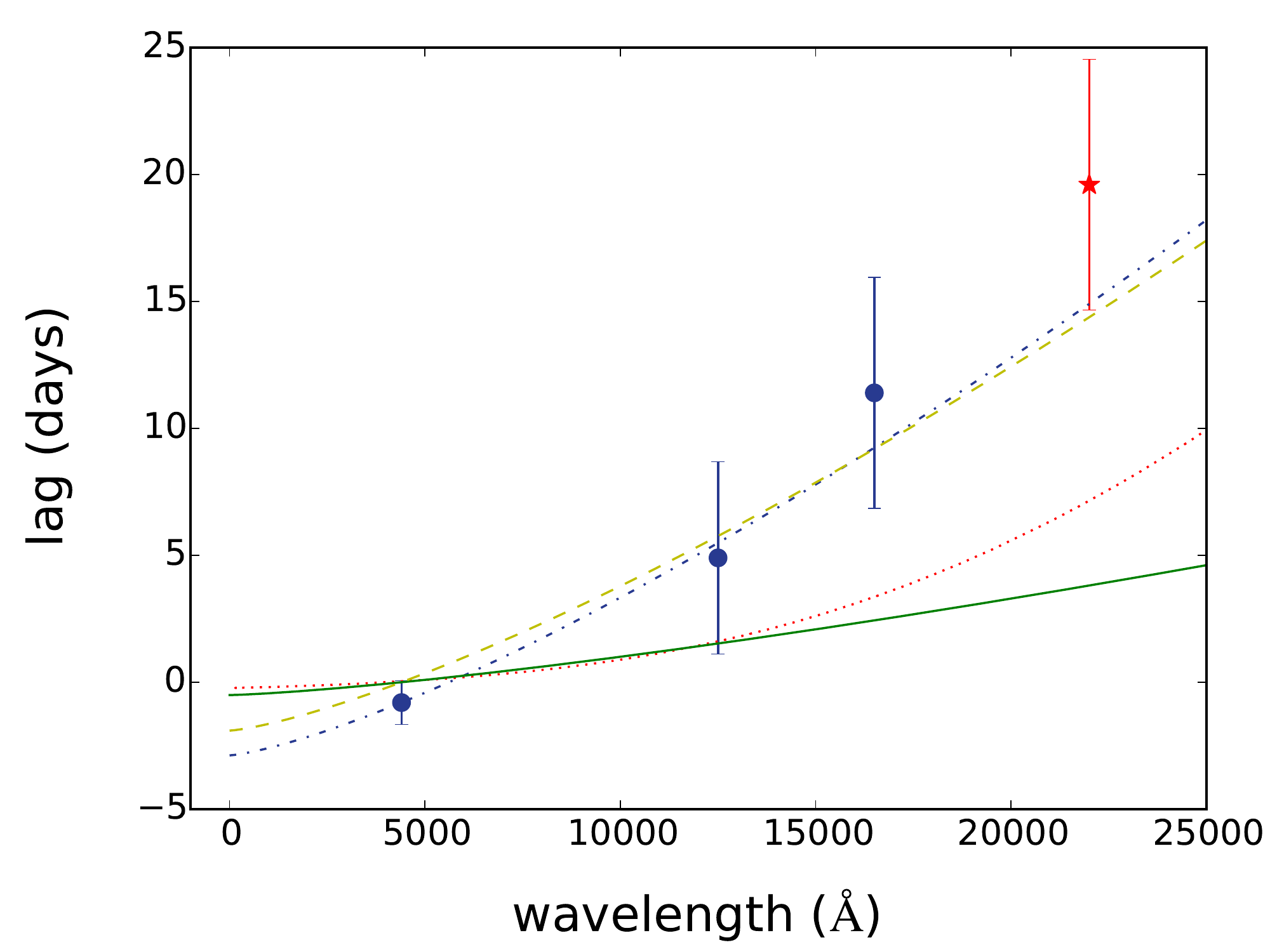}
\caption{Observed and model lags as a function of wavelength. V, J and
  H 'disc' lags were determined from the JAVELIN probability
  distributions and are shown as blue circles; the K-band 'torus' lag
  is shown with a red star (see text for details). The blue dot-dashed
  line corresponds to the best fit assuming the functional form $\tau
  = A ((\lambda/\lambda_0)^{4/3}-1)+B$. The yellow dashed line
  corresponds to the best fit for a standard optically thick
  geometrically thin accretion disk model of the form $\tau = 3\!
  \times\! 10^{-10} \lambda^{4/3} \dot{m}^{1/3} M^{2/3}$, where
  $\dot{m}$ and $M$ are the accretion rate and black hole mass, which
  are left as free parameters. The best estimates of $\dot{m}$ and $M$
  give the green solid line, which clearly under-predicts the observed
  lags. The red dotted line includes the effects of X-ray heating for
  a flared disc, as described in more detail in the text.}
\label{ad_profile}
\end{figure}

In Figure \ref{ad_profile} we present the JAVELIN 'disc' lags
determined as given above (blue circles), as a function of
wavelength. The K-band JAVELIN 'torus' lag is also included (red
star). To test whether we are in fact seeing disc emission in the
JAVELIN 'disc' lags, we can look at the correlation of the lags with
wavelength and see whether these are in agreement with the predicted
relation $\tau \propto \lambda^{4/3}$ for the outward light travel
time in an optically thick geometrically thin accretion disc.

We determined a weighted fit to the JAVELIN V, H and J 'disc'
lags. Following Edelson et al.~(2015), first we fit a function of the
form $\tau = A ((\lambda/\lambda_0)^{4/3}\!- 1)\! +B$, with
$\lambda_0=4000$\AA. The best fit requires $A=2.4$ and $B=-0.9$ and is
presented as a blue dashed-dotted line in Figure \ref{ad_profile}.

A fit setting $B\!=\!0$ was done in a physically motivated way, as
follows. We included the prescription for the light travel time
across a standard optically thick geometrically thin accretion disk
model of the form $\tau = 3\! \times\! 10^{-10} \lambda^{4/3}
\dot{m}^{1/3} M^{2/3}$, for $\tau$ expressed in days and $\lambda$ in
\AA, and where $\dot{m}$ is the accretion rate in Eddington units and
$M$ is the black hole mass in units of solar masses (e.g., Shakura \&
Sunyaev 1973). The green solid line in Figure \ref{ad_profile} shows
the predicted lags for $M=5 \times 10^6$ M$_{\sun}$ (McHardy et
al.~2005) and $\dot{m}=0.04$ (see next section). Clearly, the lags are
up to a factor 4 longer than predicted by the accretion model. A fit
to the data leaving $\dot{m}$ and $M$ as free parameters yields
$\dot{m}=0.2$ and $M=1.1 \times 10^7$ M$_{\sun}$, both much larger
than the estimates, and is shown in Figure \ref{ad_profile} with a
yellow long dashed line.

Illumination or heating of the disc by the central X-ray emission can
modify the disc temperature profile. Since the disc becomes hotter, a
region emitting most of the radiation corresponding to a given
wavelength will move outwards to larger radii. For a flared disc, the
changes can be even more significant as a larger fraction of the X-ray
flux is intercepted by the disc. Besides, it is easy to argue that it
is just consistent to include heating from the central X-ray
source\footnote{This is done in the 'lamp-post' approach. Whether disc
  UV and optical self-illumination is also a contribution is not
  clear, although only a very flared geometry would allow this to be a
  meaningful contributor to the general heating.} when computing the
$\tau \propto \lambda^{4/3}$ correlation: the cross correlation
analysis clearly demonstrate that the observed lags are consistent
with the light travel time across the system and that such a signal is
propagating outwards; in other words, we see the disc being
illuminated from the centre.

We follow Lira et al.~(2011) and assume $M=5 \times 10^6$ M$_{\sun}$,
$\dot{m}=0.04$, a 2-10 keV X-ray power of $7.0 \times 10^{42}$ ergs/s
(for a mean X-ray flux of $\sim 4 \times 10^{-11}$ ergs/s/cm$^{2}$, as
seen in our observations), a factor 2 to take the 2-10 keV X-ray
luminosity to the the full 0.01-500 keV range, a disc innermost stable
orbit at 3 $R_g$, a low disc albedo of 10\%, a heavily flared disc
with a power law profile of index 1.5 and characteristic radius $R_0$
of 100 $R_g$ (i.e., a disc height of the form $H \propto
(R/R_0)^{1.5}$), and a height of the X-ray source above the disc of 10
$R_g$. This predicts unreddened V-band fluxes $f_{V {\rm Grav}} = 1.6
\times 10^{-14}$ and $f_{V {\rm XR}} = 4.1 \times 10^{-14}$
ergs/s/cm$^2$/\AA\ due to the release of gravitational and X-ray
heating, respectively. After an extinction of $E(B\!-\!V)=0.6$ is
applied, this corresponds to $2.9 \times 10^{-15}$ and $7.4 \times
10^{-15}$ ergs/s/cm$^2$/\AA. Clearly, the predicted V-band flux due to
X-ray heating is too large to account for our observations. If we take
into account our (rather extreme) estimate of the host contribution to
the V-band (see Section 3.1), then $f_{V \star} + f_{V {\rm Grav}} =
5.6 \times 10^{-15}$ erg/s/cm$^2$/\AA, about 35\%\ above the observed
mean flux value in the V-band light curve.

The predicted lags as a function of wavelength for the described disc
model is plotted in Figure \ref{ad_profile} with a dotted red line. As
can be seen, even though the predicted lags are significantly
increased at longer wavelengths, this effect is not enough to account
for the discrepancies with the observations. Only with a X-ray power 4
times the observed value it is possible to reproduce the observed
lags. This is not only energetically impossible but it would also give
V-band fluxes in huge disagreement with our observations.

These results could be interpreted as torus emission still dominating
the response of the 'disc' near-IR peaks observed in the JAVELIN
probability distributions. In fact, the JAVELIN K-torus-lag (red star
in Figure \ref{ad_profile}) seems to nicely follow the trend of the
'disc' lags determined at lower wavelengths. Alternatively, our
theoretical prescription might not be correct.

A pattern where lags are found to be longer than predicted, has
recently been seen in the UV and optical lags of NGC~5548 (McHardy et
al.~2014; Edelson et al.~2015) and for the UV to the near-IR in
NGC~2617 (Shappee et al.~2014), while on the other hand the model
nicely fits the observed NGC~4395 lag values (McHardy et al., in
prep). The disagreement with the predictions from an optically thick
geometrically thin accretion disc seems to suggest that accretion
discs are larger than predicted by the theory, in line with the
results from microlensing of distant quasars (Mosquera et
al.~2013). On the other hand, general accretion disc models are able
to successfully reproduce the rest-frame UV to optical continuum of
quasars at $z\sim 1.55$, as shown by Capellupo et al.~(2015). Hence,
it seems it is still early days to draw firm conclusions on the
validity of the current accretion disc models to prove or disprove
these different lines of evidence.

\subsection{The dust sublimation radius and the bolometric luminosity of MCG-6-30-15}

We have no direct indication for the presence of a dusty torus in
MCG-6-30-15. However, a few lines of evidence suggest that indeed, hot
dust is found beyond the location of the accretion disc: 1) the large
power at low frequencies seen in the near-IR power spectral density;
2) the red slope between the H and K band photometric measurements; 3)
the structure in the cross correlation DCF and JAVELIN results of the
J and H bands, which hint at the presence of more than one reprocessor
in MCG-6-30-15.

Assuming that the inner face of the torus is located at 20 light-days
from the central source, which corresponds to the correlated signal
seen between the K-band and the B-band, and using the relation between
the source luminosity and the sublimation dust radius (see, e.g.,
Barvainis (1987); Nenkova et al.~(2008)), we find that the bolometric
luminosity for MCG-6-30-15, $L_{bol}$, is $\sim 2 \times 10^{42}$
ergs/s, for a sublimation temperature of 1500 K.

However, the luminosity--sublimation-radius expression is found to
systematically overestimate the torus sizes as determined by dust
reverberation measurements (Kishimoto et al.~2007; Koshida et al.,
2014). Hence, a more accurate luminosity estimate would be given by
the empiric relation determined from the reverberation analysis of 17
nearby Seyfert galaxies by Koshida et al.~(2014): $\log \tau = -0.2\,
M_V - 2.1$ (light-days), where $\tau$ is the lag between V and the
K-band. A K-band lag of 20 days for MCG-6-30-15 then corresponds to an
intrinsic nuclear absolute V-band magnitude of $-17.0$, or a
luminosity of $2.8 \times 10^{38}$ ergs/s/\AA. Adopting a $f_{\lambda}
\propto \lambda^{-1.56}$ slope for the optical continuum (vanden Berk
et al.~2001), and a luminosity dependent bolometric correction
(Marconi et al.~2004), we find a bolometric luminosity of $3 \times
10^{43}$ ergs/s. For a $M=5 \times 10^6$ M$_{\sun}$ this translates
into an Eddington ratio of 0.04.

Reynolds et al.~1997 estimated a total luminosity for MCG-6-30-15 of
$L_{bol} = 8 \times 10^{43}$ ergs/s from the direct integration of the
observed SED after assuming an extinction of $E(B\!-\!V)=0.6$.
Correcting for the double counting of the IR emission, which
corresponds to optical and UV emission absorbed by the dusty torus and
reprocessed into our line of sight, and the used cosmology (Ho=50
km/s/Mpc, Reynolds, private communication), then $L_{bol} \sim 2
\times 10^{43}$ ergs/s, in good agreement with our findings. This is
somewhat below the values derived by Vasudevan et al.~(2010) based on
the combined analysis of hard X-rays (14-195 keV) and IRAS photometry,
who found $L_{bol} = 4-8 \times 10^{43}$ ergs/s. The discrepancy is
most likely due to host contamination of the IRAS measurements, as
clearly discussed by Vasudevan et al.~(2010).

We can make some consistency checks using our observed light curves
and the previous accretion disc modelling. At 37 Mpc of distance, the
above derived V-band nuclear intrinsic luminosity corresponds to a
flux $f_{V {\rm RM}} = 1.7 \times 10^{-15}$ erg/s/cm$^2$/\AA\ (RM as in
reverberation mapping). This is about an order of magnitude below the
predicted V-band flux due to the release of gravitational energy,
$f_{V {\rm Grav}}$, which is basically dependent on the adopted values
for the black hole mass and accretion rate only (with the value of the
innermost orbit having a minor role as the V-band emission comes from
radii located further out in the disc). Then, for these two results to
match, either the black hole mass or the accretion rate would have to
be scaled down significantly.

On the other hand, after applying an extinction value of $E(B\!-\!V) =
0.6$ to $f_{V {\rm RM}}$, we determine an observed nuclear flux of
$\sim 4 \times 10^{-16}$ erg/s/cm$^2$/\AA. This is less than the
peak-to-peak variation seen in the V-band light curve, which
corresponds to $\sim 7 \times 10^{-16}$ erg/s/cm$^2$/\AA, and of
course it is solely due to the active nucleus. 

In summary, accretion theory seems to over-predict the observed V-band
flux level, while dust reverberation seems to under-predict it. Given
the many uncertainties like the host contribution to the observed
light curves, the level of obscuration towards the nucleus of
MCG-6-30-15, its black hole mass and accretion rate, it is not totally
surprising to find conflicting results.

\section{Summary}

We present long term monitoring of MCG-6-30-15 in X-rays, optical and
near-IR wavelengths, collected over five years of observations. We
determined the power spectral density of all the observed bands and
find that the host contribution needs to be taken into account to
obtain reasonable results. The lag determined between the X-ray Q flux
and the optical bands is consistent with zero days, while the lags
between optical and near-IR bands correspond to values in the 10 to 20
day range. Filtering the light curves in frequency space shows that
most of the correlation is due to the fastest variability. We discuss
the nature of the X-ray variability and argue that this must be
intrinsic and cannot be accounted for by a absorption episodes due to
material intervening in the line of sight. It is also found that the
lags agree with the relation $\tau \propto \lambda^{4/3}$, as expected
for an optically thick geometrically thin accretion disc, although for
a larger disc than that predicted by the measured black hole mass and
accretion rate in MCG-6-30-15. We find some evidence for a truncation
of the disc at a distance of 15 light-days. Indirect evidence suggests
that the torus might located at $\sim 20\ R_g$ from the central
source. This implies an AGN bolometric luminosity of $\sim 3 \times
10^{43}$ ergs/s/cm$^2$.

\section*{Acknowledgments}

PL and PA are grateful of support by Fondecyt projects 1120328 and
1140304, respectively.

\section*{References}

Ar\'evalo, P., Papadakis, I., Kuhlbrodt, B., et al., 2005, A\&A, 430, 435\\
Ar\'evalo, P., Uttley, P., Kaspi et al., 2008, MNRAS, 389, 1479\\
Ar\'evalo, P., Uttley, P., Lira, et al., 2009, MNRAS, 397, 2004\\
Ar\'evalo, P., Churazov, E., Zhuravleva, I., et al., 2012, MNRAS, 426, 1793\\
Barvainis, R., 1987, ApJ, 320, 537\\
Breedt, E., McHardy, I. M., Ar\'evalo, et al., 2010, MNRAS, 403, 605\\
Cameron, D. T., McHardy, I., Dwelly, T., et al., 2012, MNRAS, 422, 902\\
Capellupo, D. M., Netzer, H., Lira, P., et al., 2015, MNRAS, 446, 3427\\
Chiang, Chia-Ying, Fabian, A. C., 2011, MNRAS, 414, 2345\\
Done, Chris, Gierlinski, Marek 2005, MNRAS, 364, 208\\
Edelson, R. A., Krolik, J. H., 1988, ApJ, 333, 646\\
Edelson et al.~2015, ApJ, submitted\\
Emmanoulopoulos, D., McHardy, I. M., Papadakis, I. E., 2011, MNRAS, 416, L94\\
Fabian, A. C., Vaughan, S., Nandra, K., et al., 2002, MNRAS, 335, L1\\
Guainazzi, M., Matt, G., Molendi, S., Orr, A., 1999, A\&A, 341, L27\\
Inoue, Hajime, Matsumoto, Chiho 2003, PASJ, 55, 625\\
Iwasawa, K., Fabian, A. C., Reynolds, C. S., 1996, MNRAS, 282, 1038\\
Kara, E., Fabian, A. C., Marinucci, et al., 2014, MNRAS, 445, 56\\
Kelly, Brandon C., Bechtold, Jill, Siemiginowska, Aneta, 2009, ApJ, 698, 895\\
Lee, J. C., Fabian, A. C., Brandt, W. N., 1999, MNRAS, 310, 973\\
Lira, P., Ar\'evalo, P., Uttley et al., 2011, MNRAS, 415, 1290\\
Ludlam, R. M., Cackett, E. M., G\"ultekin, K., 2015, MNRAS, 447, 2112\\
Marinucci, A., Matt, G., Miniutti, et al., 2014, ApJ, 787, 83\\
Matsumoto, Chiho, Inoue, Hajime, Fabian, Andrew C., et al., 2003, PASJ, 55, 615\\
McHardy, I. M., Gunn, K. F., Uttley, P., et al., 2005, MNRAS, 359, 1469\\
McHardy, I. M., Cameron, D. T., Dwelly, et al., 2014, MNRAS, 444, 1469\\
Miller, L., Turner, T. J., Reeves, J. N. 2008, A\&A, 483, 437\\
Miller, L., Turner, T. J., Reeves, J. N. 2009, MNRAS, 399, L69\\
Mosquera, A.~M., Kochanek, C.~S., Chen, B., Dai, X., Blackburne, J.~A., Chartas, G., 2013, ApJ, 769, 53\\
Nandra, K., Pounds, K. A., Stewart, G. C., 1990, MNRAS, 242, 660\\
Nenkova, M., Sirocky, M. M.; Nikutta, R., Ivezic, Z., Elitzur, M., 2008, ApJ, 685, 160 \\
Peterson, B. M., Ferrarese, L., Gilbert, K. M., et al., 2004, ApJ, 613, 682\\
Pounds, K. A., Turner, T. J., Warwick, R. S., 1986, MNRAS, 221 P7\\
Raimundo, S. I., Davies, R. I., Gandhi, P., et al., 2013 MNRAS, 431, 2294\\
Reynolds, C. S., Fabian, A. C., Nandra, K., et al., 1995 MNRAS, 277, 901\\
Reynolds, C. S., Ward, M. J., Fabian, A. C., et al., 1997, MNRAS, 291, 403\\
Smith, R.; Vaughan, S., 2007, MNRAS, 375, 1479\\
Suganuma, Masahiro, Yoshii, Yuzuru, Kobayashi, Yukiyasu, et al., 2006, ApJ, 639, 46\\
Shappee, B. J., Prieto, J. L., Grupe, D., et al., 2014, ApJ, 788, 48\\
Tanaka, Y., Nandra, K., Fabian, A. C.,  1995, Nature, 375, 659\\
Timmer, J., Koenig, M. 1995, A\&A, 300, 707\\
Uttley, Philip, McHardy, Ian M. 2001, MNRAS, 323, L26\\
Uttley, P., McHardy, I. M., Vaughan, S., 2005, MNRAS, 359, 345\\
Uttley, P. 2007 in 'The Central Engine of Active Galactic Nuclei', Ed L. C. Ho and J-M Wang,\\
Vasudevan, Fabian, Gandhi, Winter, Mushotzky, 2010, MNRAS, 402, 1081\\
Vaughan, S., Fabian, A. C., Nandra, K., 2003, MNRAS, 339, 1237\\
Zu, Ying, Kochanek, C. S., Kozlowski, Szymon, Udalski, Andrzej, 2013, ApJ, 765, 106\\
Zu, Ying, Kochanek, C. S., Peterson, Bradley M., 2011, ApJ, 735, 80\\

\end{document}